\documentclass[nocopyrightspace,fleqn,10pt]{sigplanconf}

\usepackage{amsmath}

\usepackage{datetime}

\usepackage{subfigure}
\usepackage[table]{xcolor}
\usepackage{xspace}

\usepackage[T1]{fontenc}
\usepackage{inconsolata}
\definecolor{bluekeywords}{rgb}{0.13,0.13,1}
\definecolor{greencomments}{rgb}{0,0.5,0}
\definecolor{redstrings}{rgb}{0.9,0,0}

\usepackage{listings}
\newcommand{\pcie}{PCIe\xspace} \newcommand{\Gbps}{Gb/s\xspace} \newcommand{\musec}{$\mu$s\xspace}

\newboolean{comments}
\setboolean{comments}{true}

\newcommand{\csharp}{C$^\#$\xspace}

\usepackage{amssymb} \usepackage{hyperref}
\usepackage{graphicx}
\usepackage{rotating}
\usepackage{float}
\usepackage{booktabs}

\usepackage[english]{babel}
\usepackage{mathpartir}
\usepackage{adjustbox}

\usepackage{stmaryrd}
\usepackage{amsmath}
\newcommand{\fatbrack}[1]{\ensuremath{\left\llbracket #1 \right\rrbracket}}

\usepackage{amsthm}
\theoremstyle{definition}
\newtheorem{defn}{Definition}[section]

\newcommand{\somecode}[1]{\ensuremath{\ulcorner #1 \urcorner}}
 
\makeatletter
\begin{document}

\setlength{\pdfpageheight}{\paperheight}
\setlength{\pdfpagewidth}{\paperwidth}

\conferenceinfo{CONF 'yy}{Month d--d, 20yy, City, ST, Country}
\copyrightyear{20yy}
\copyrightdata{978-1-nnnn-nnnn-n/yy/mm}
\copyrightdoi{nnnnnnn.nnnnnnn}

\titlebanner{DRAFT: please do not circulate}        \preprintfooter{Generated at: \currenttime}   
\title{Extending programs with debug-related features, with application to hardware development}

\newcommand{\cucl}[1]{$\text{#1}^\flat$}
\newcommand{\icl}[1]{$\text{#1}^\natural$}
\newcommand{\qmul}[1]{$\text{#1}^\sharp$}

\newcommand{\spejs}{\hspace{1.5mm}}

\authorinfo{\begin{minipage}{0.95\textwidth}\centering\cucl{Nik Sultana} \spejs
  \cucl{Salvator Galea} \spejs \cucl{David Greaves} \spejs \cucl{Marcin
  W\'ojcik} \spejs \cucl{Noa Zilberman} \spejs \qmul{Richard Clegg}
  \icl{Luo Mai} \spejs\cucl{Richard Mortier} \spejs \icl{Peter Pietzuch} \spejs \cucl{Jon Crowcroft} \spejs \cucl{Andrew W Moore}
\end{minipage}}
{${}^\flat$Cambridge University \and ${}^\natural$Imperial College, London
\and ${}^\sharp$Queen Mary University, London}{}

\maketitle

\begin{abstract}
The capacity and programmability of reconfigurable hardware such as FPGAs has
improved steadily over the years, but they do not readily provide any
mechanisms for monitoring or debugging running programs.  Such mechanisms need
to be \emph{written into} the program itself. This is done using \emph{ad hoc}
methods and primitive tools when compared to CPU programming. This complicates
the programming and debugging of reconfigurable hardware.

We introduce \emph{Program-hosted Directability} (PhD), the extension of
programs to interpret \emph{direction commands} at runtime to enable debugging,
monitoring and profiling.
Normally in hardware
development such features are fixed at compile time.
We present a language of directing commands, specify its semantics in terms of
a simple controller that is embedded with programs,
and implement a prototype for directing network programs running in hardware.
We show that this approach affords significant flexibility with low impact on
hardware utilisation and performance.
\end{abstract}


\section{Introduction}
When debugging and monitoring programs running on microprocessors we usually
benefit from hardware support that is leveraged by an Operating System to
inspect and modify running processes~\cite{4218560}. But when a program is run
on \emph{reconfigurable hardware} platforms, one does not usually have an
operating system, a notion of process, nor any hardware support for
debugging.
Programs must contain additional logic to enable \emph{debugging},
\emph{monitoring}, and \emph{profiling} during their execution, because
the environment does not provide visibility into running programs by default.

Field-Programmable Gate Array devices (FPGAs) are a form of programmable
hardware consisting of a grid of logic blocks whose function and wiring can be flexibly
reconfigured. FPGAs are used to perform functions for which a
full-featured general-purpose CPU is not appropriate. For such functions, FPGAs can operate at a
higher throughput and consume much less electricity than CPUs~\cite[\S4.4]{mittal2014survey}.
This makes FPGAs especially appealing for some problems and environments, a
recent example being datacentres~\cite{6853195}.

Despite their appeal as a computing device, the programmability of FPGAs has been
hampered by the need for low-level hardware-description languages
traditionally used to program them, such as Verilog and VHDL, which
``requires programmers to assume the role of hardware
designers''~\cite{So:2002:CAF:512529.512550}.
Research has yielded various approaches for \emph{high-level
synthesis} (HLS): this tends to consist of using (fragments of)
existing languages, such as C or Java, to describe hardware.

In addition to a \emph{programmability} gap between FPGAs and CPUs, there is a
\emph{debuggability} gap that has received far less attention.
The programmability of FPGAs has improved over the years, but they are not
\emph{debuggable} by default~\cite{Potkonjak:1995:DAS:224841.225054}.
FPGAs provide no visibility into
the running program, and their standard tooling provides very limited
support for this. For full visibility one could \emph{simulate} the program, (e.g., on a workstation) but the simulation can be slower by a factor of
$10^6$~\cite{Camera:2005:IDE:1085130.1085145} because of the sheer amount of
detail that must be simulated.

In this paper we propose to improve the debuggability of programs running on
FPGAs by using a domain-specific control language inspired by \emph{program
directing}~\cite{Sosic:1992:DTP:143103.143110} to generate in-program support
for debugging.
In addition to debugging, this can be used to monitor and profile programs.

We call our approach \emph{Program-hosted Directability} (PhD).  It involves
extending the user's program to service \emph{direction commands} at runtime.
Extending the program involves inserting (i)~named \emph{extension points}
which can contain runtime-modifiable code in a computationally weak language
(no recursion), and (ii)~state to be used for book-keeping by that code, to
implement direction features.

\begin{figure}
  \centering
  \includegraphics[width=0.47\textwidth]{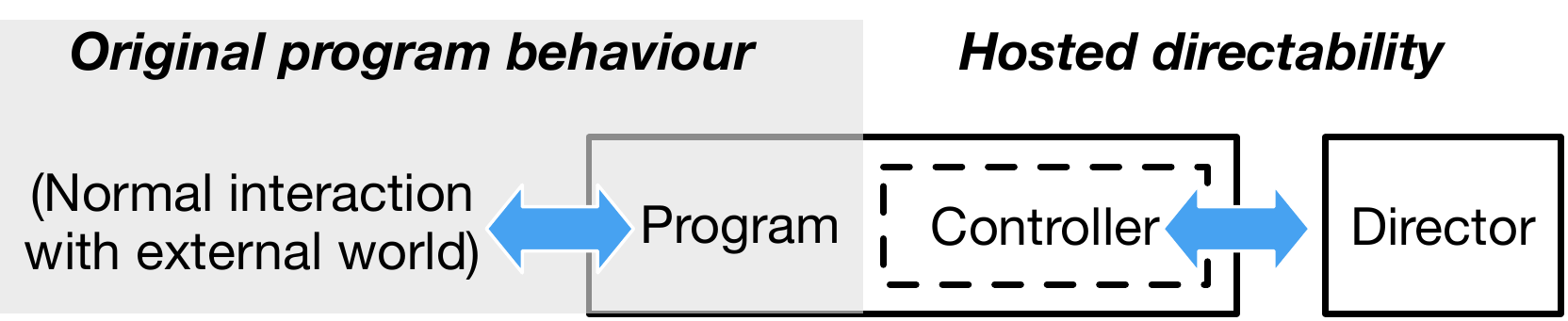}
  \caption{\label{fig:director}
  A \emph{controller} is embedded into the program, and acts as the agent of the
  \emph{director}.     The director and controller implement a protocol to exchange commands and
  their outputs.
    }
  \vspace{-2em}
\end{figure}

For example, the direction command ``$\mathsf{trace} \; X \;
\\\mathrm{max\_trace\_idx}$'' (where $X$
is a variable in the program's source code) logs updates made to a variable, and
appends these updates to a buffer (of size ``$\mathrm{max\_trace\_idx}$'' for later
inspection). We translate this command into (i)~a snippet that gets injected into
the program~(Figure~\ref{example:tracing:1})
and (ii)~a modification of the program to allocate space for this snippet,
and for state require by this snippet: such as the variable \texttt{V\_trace\_idx},
and array \texttt{V\_trace\_buf}.
More examples are given in~\S\ref{sec:emu:directing}, and the overall scheme is
shown in Figure~\ref{fig:director}.

Idiomatic direction features (such as tracing, breakpoints, etc) are compiled down
to a weak language and executed by an interpreter embedded in the program.
For consistency with the description of the interpreter's behaviour as hardware,
we refer to it as a \emph{controller}.
The controller is invoked by extension points, which are inserted into the
program through transformation. In the above example, the extension
points would be added after each \emph{update} to a variable, to ensure that the
update is considered for logging.

Extension points are added at compile time depending on what direction commands
we want to support at runtime.
For example, starting with the sequence of statements
$\ldots;s_i;s_{i+1};\ldots$ we insert two extension points $@L$ and $@M$,
resulting in the program $\ldots;s_i;@L;s_{i+1};@M;\ldots$.
When extension point $L$ is reached at runtime, a stored procedure associated
with $L$ is executed by the controller. This procedure ultimately hands back control
to the host program, or starts an interactive session with a \emph{program director},
which can send the controller further commands, and update its stored procedures.

\begin{figure}
\begin{lstlisting}[  basicstyle=\small\ttfamily,
    escapeinside={@}{@}]
if V_trace_idx < max_trace_idx then
  V_trace_buf[V_trace_idx] := V;
  inc V_trace_idx;
  continue
else
  inc V_trace_overflow;
  break
\end{lstlisting}
  \caption{\label{example:tracing:1}
Code that implements the direction command ``$\mathsf{trace} \; X \;
\mathrm{max\_trace\_idx}$''. If the buffer is not full then the new value of
X is logged, the index incremented, and control is handed back to the program that hosts this code.
Otherwise indicate depletion of the associated buffer resource
and break the program's execution. }  \vspace{-2em}
\end{figure}

Our work systematises
\emph{ad hoc} debugging and profiling extensions to programs, and generalises the
facilities currently made available for hardware development. Moreover, we can
include only the features needed, thus improving the utilisation of the
hardware, and its power consumption. The approach is extensible: one can code
additional direction commands, or variations~(\S\ref{sec:casp:examples}).
In this paper we show how to give semantics to program direction commands in
terms of the placement of extension points, the code that is to be run by the
controller, and the interaction between the controller and the direction tool
that manages it.

We prototype PhD using an HLS and obtain a uniform interface for directing the
software and hardware instances of the same program, allowing us to unify the
debugging of these instances (which otherwise require diverse tools).

We believe that PhD can yield practical benefits in hardware development and deployment.  As a
technique, PhD is vendor-neutral and compiler-neutral, and the communication
between the director and controller can be adapted to suit the program. For
example, in our prototype we send direction commands via the network. PhD can
be implemented in different ways: in our prototype we implemented it entirely
in a high-level language via HLS, but one could also bolt it onto more finely
engineered hardware blocks.  Despite running as hardware, the directability
features can be enabled and reconfigured at runtime: this consists of updating
the code stored by the controller. In contrast, existing techniques for FPGAs
involve including \emph{fixed function} circuits in the design. Finally, PhD
extends a program with a \emph{direction mode}, thus facilitating the in-field
debugging of programs.

Through PhD we hope to contribute to the convergeance between the debuggability
of programs on FPGAs, and those on CPUs, which enjoy extensive hardware and OS
support, and which in turn benefits sophisticated monitoring systems such as
DTrace~\cite{gregg2011dtrace}.
The ideas described in this paper are not necessarily tied to the languages,
compilers, FPGA or other equipment we used in our prototype.
We make the following contributions:
\begin{itemize}

  \item We describe how to translate familiar high-level idioms for debugging,
    profiling and monitoring~(\S\ref{lang:direct}), which we call
    \emph{direction commands}, into a low-level language for controlling
    program state at runtime~(\S\ref{casp-machines}), for an example language~\S\ref{lang:prog}.

  \item We relate the direction commands with our low-level language through a
    specification~(\S\ref{directability-ordering}) in which programs are ordered
    by the directability features they support.  This provides a
    basis on which we can reason that one program is ``more debuggable'' than
    another at runtime.

  \item A prototype implementation and its evaluation~(\S\ref{sec:eval}), where
    we measure the effect of directability on FPGA utilisation and performance.
\end{itemize}

\section{FPGA debugging gap}
\label{sec:debugginggap}
Irrespective of how we write our program as a hardware description, once a
so-called ``bitstream'' is generated that configures an FPGA to run our program,
the program can only be tested as a black box, and we cannot understand or
further influence its behaviour at that stage.

The need for better FPGA debugging features is becoming more urgent since:
\begin{itemize}
\item FPGA chips are getting larger, which allows them to run more complex
  programs. Complex programs are more likely to be buggy, which necessitates
    more debugging.
\item FPGAs are also used for simulating hardware designs, since
behavioural simulations are very slow.~\cite{Wang:2011:DPA:1999946.1999970,7294021}
\item FPGAs are being deployed in large production environments, such as
  Microsoft's datacentres~\cite{large-scale-reconfigurable-computing-microsoft-datacenter}.
\end{itemize}

Hardware development has driven the development of formal methods to establish
system correctness~\cite{Fix2008}, which enabled the development for methods
for software~\cite{Ball:2006:TSA:1218063.1217943,Godefroid:2012:SWF:2093548.2093564}.
Unfortunately the verification is done on the Register-Transfer Level (RTL), a
higher-level description of a hardware circuit in languages such as Verilog, and
not on the generated bitstream. Thus debugging may still be needed.

\subsection{Debugging(FPGA) != Debugging(CPU)}
Debugging concepts from software do not
correspond directly with debugging the same program on an FPGA.
A lot of the core issues involved were discussed in the pioneering work of
\citet{Koch:1998:BBD:290833.290843} and in many other work on HLS
debugging~\cite{6927498,Monson:2015:UST:2684746.2689087}.
These are the main points:
\begin{itemize}
  \item Multiple source lines might be executed concurrently in hardware.  Code
    is represented at the source, register-transfer, and gate levels. This has
    important consequences for debugging, described in the following points. The correspondence
    between these levels necessitates keeping metadata from compilation for
    debugging.
  \item Depending on the debugged artefact, \emph{stepping} by (source) line
    might be less useful than stepping by cycle.
  \item \emph{Breakpoints} become more tricky to interpret, since they are
    usually set on a specific line of code. In hardware there may be several overlapping lines being executed
    at a breakpoint~\cite[\S4.3]{Koch:1998:BBD:290833.290843}. Moreover, the
    output of operations on previous lines might only be available \emph{after} some clock
    cycles have elapsed. Depending on what the user wishes to do, they might
    prefer if the breakpoint is triggered after the elapsing of these cycles.
    Furthermore, part of the next line of source code might have started
    executing. This suggests that a strict indication of sequentiality needs to
    be communicated to the HLS compiler if the usual breakpoint semantics are
    desired.
  \item As mentioned in the introduction, FPGAs do not provide hardware support
    such as debug registers to assist with analysing running programs.
\end{itemize}

\subsection{Current techniques for FPGA debugging}
Some existing techniques help narrow the debugging gap on FPGAs.
\emph{Co-simulation} involves comparing the behavioural simulation between HLS
and RTL. This can be considered a special case of \emph{relative
debugging}~\cite{sosivc1997guard} but it does not provide visibility into the
hardware instance of a program.
Another technique involves \emph{in-system} testing: testing a large part of the
system, though possibly not all. This does not provide visibility into the
hardware either.

Current practice employs two techniques for FPGA debugging.
\textbf{Trace buffers} are the most popular technique for
debugging FPGAs. It requires a programmer to identify signals of interest in the circuit at
compile time, then an \emph{embedded logic analyser} is synthesised that
uses on-chip memory to record traces for these signals.  This suffers from
two problems: only a limited number of signals may be viewed (limited by on-chip
memory), and traces have a limited window size (for the same reason).
Traces may be conditional, to avoid using up buffer space unnecessarily,
but this technique is difficult to use because it involves generating the bitstream
each time.
\textbf{Register scanning}  allows you to see the values of all registers on the FPGA, but
requires ``stopping the world'' to enable reading and sending it off-chip.
This slows down tests, and thus register scanning has been supplanted by trace
buffers. 

Both register scanning and trace buffers usually send recorded data off-chip
via the JTAG (Joint Test Action Group) interface, a standardised
instrumentation method~\cite{Bennetts1991}. This method is not scalable, since
its transfer rate is far less than the FPGA throughput.

In summary, existing techniques consist of including ``fixed function'' modules
as part of your hardware description at compile time. This has the advantage of
being lightweight since these circuits are specialised to perform a single
function, but it has the disadvantage of being inflexible. Generating a hardware
bitstream can take hours, and the added overhead costs for
runtime-reconfigurable debuggability and monitoring features might not be
affordable in some use-cases. Furthermore, these techniques cannot be used in
production environments.

\subsection{High-Level Synthesis (HLS)}
HLS involves the use of a high-level language, such as Java, C, C++, and OCaml,
to write a hardware description. This takes advantage of the features of, and
tooling available for, the high-level language. An RTL description is then
generated from the high-level description.

Using an HLS to describe hardware enables one to run the HLS description as
a software program, and to debug it as such, by using standard tools to compile
and debug Java programs for instance. This prunes bugs from the eventual bitstream
and avoids regenerating the bitstream.
Irrespective of whether testing is directed at software or hardware, it can take
many tests to find a fault. 
Software-based testing could help detect logic errors in our code, but it could
not help us find some important classes of problems: `[Testing in] the silicon
mode permits the analysis of bugs that are ``invisible'' at the RTL
level'~\cite{6927496}. We outline the main cases below:
\begin{enumerate}
  \item \emph{Interface mismatch.}
    \label{point:interface-mismatch}
    We need to understand whether a problem occurs because of a mismatch between
    one module and the rest of the circuit. Recall that
    behavioural simulation might not be applied to the whole design, and
    incorrect assumptions about the enclosing circuit can result in the
    simulation test succeeding but the hardware tests failing.

  \item \emph{Reproducability.}
   Some faults are triggered during high-throughput tests, and are difficult to
   find when testing other instances of the program. Other faults result from
   features of the hardware and transient environmental states -- such as
    ``Single Event Effects'' manifested through the flipped or stuck bits from
    the interaction of charged particles with semiconductors~\cite{Sari:2014:SEV:2554688.2554767,Krishnaswamy:2008:RTM:1391469.1391703}.

  \item \emph{Toolchain problems.}
    Diagnosing bugs in the compiler toolchain becomes easier if we can see into
    the compiled program's operation.      Bugs are not unusual in both HLS and RTL toolchains.
\end{enumerate}

\subsection{Current research on FPGA debugging}
Various improvements have been explored for the techniques described above. For
example, one could write summaries to the trace-buffer, rather than the explicit
trace~\cite{6927498}. Another idea consists of multiplexing all signals and
choosing which to observe at debug time rather than
at compile time~\cite{Hung:2013:TSO:2435264.2435272}, enabling ``observation
without recompilation''.  An additional idea is to buffer into a fast external
memory~\cite{7294023}.

\newcommand{\mcrot}[4]{\multicolumn{#1}{#2}{\rlap{\rotatebox{#3}{#4}~}}}

\newcommand{\rotcheckmark}{\small \checkmark}
\newcommand{\rotNothing}{\small }
\newcommand{\rotN}{\small N}
\newcommand{\rotC}{\small C}
\newcommand{\rotS}{\small S}
\newcommand{\rotH}{\small H}

\begin{table*}
  \begin{center}
  \begin{tabular}{rcccccccccccccc||cccccc}
    \bf{System} \kern-1em \rotatebox{90}{\kern1.5em \bf{Features}}

    & \mcrot{1}{l}{90}{state inspection}
    & \mcrot{1}{l}{90}{trace recording}
    & \mcrot{1}{l}{90}{state updating}
    & \mcrot{1}{l}{90}{extension points}
    & \mcrot{1}{l}{90}{break points}
    & \mcrot{1}{l}{90}{stepping}
    & \mcrot{1}{l}{90}{interruption} 
        & \mcrot{1}{l}{90}{fine granularity}     & \mcrot{1}{l}{90}{assertion checking}
    & \mcrot{1}{l}{90}{hang detection}
    & \mcrot{1}{l}{90}{timing checks}

        & \mcrot{1}{l}{90}{software instance} 
    & \mcrot{1}{l}{90}{runtime reconfigurable}
        & \mcrot{1}{l}{90}{HLS (vs HDL)}

    & \mcrot{1}{l}{90}{\underline{\bf N}etwork/\underline{\bf C}ontrol}

    & \mcrot{1}{l}{90}{use leftover resource}

    & \mcrot{1}{l}{90}{embed at \underline{\bf S}ource/\underline{\bf H}DL} 

\\
\midrule
\rowcolor{black!10} \cellcolor{white}
    \cite{Sosic:1992:DTP:143103.143110} \bf{Dynascope} & \rotcheckmark & \rotcheckmark &
    \rotcheckmark & \rotcheckmark & \rotcheckmark & \rotcheckmark     & \rotcheckmark
    & \rotcheckmark & \rotcheckmark &  &  & \rotcheckmark
    & \rotcheckmark
    & \cellcolor{black} & \cellcolor{black} & \cellcolor{black} & \cellcolor{black}
     
\\
\cite{6927498} \bf{HLS-Scope}
    & \rotcheckmark & \rotcheckmark
    &  &  & \rotcheckmark & \rotcheckmark &      & \rotcheckmark &  &  &  &
    &
        & \rotcheckmark & \rotC &  & \rotS

\\
\rowcolor{black!10} \cellcolor{white}
\cite{6927496} \bf{Inspect} & \rotcheckmark &&&&\rotcheckmark&\rotcheckmark&&\rotcheckmark
  & & & & \rotcheckmark &  & \rotcheckmark & \rotC & & \rotH \\

\cite{7294023}
    & \rotcheckmark &
    &  &  &  &  &      &  &  &  &  &
    &
        & \rotcheckmark & \rotC &  & \rotS
\\

\rowcolor{black!10} \cellcolor{white}
    \cite{Hung:2014:AFD:2597648.2566668} \bf{QuickTrace} & \rotcheckmark
    &&&&&&& \rotcheckmark
  & & & & \rotcheckmark & \rotcheckmark & \rotcheckmark & \rotN & \rotcheckmark & \rotH \\

\cite{Koch:1998:BBD:290833.290843} \bf{SLE}/CADDY & \rotcheckmark &&&&\rotcheckmark&\rotcheckmark&& \rotcheckmark
  & & & & \rotcheckmark &  & \rotcheckmark & \rotC & & \rotH \\

\rowcolor{black!10} \cellcolor{white}
\cite{Monson:2015:UST:2684746.2689087}     & \rotcheckmark & \rotcheckmark
    &  &  &  &  &      &  &  &  &  &
    &
        & \rotcheckmark & \rotC &  & \rotS
\\

\cite{Curreri:2011:HSI:1972682.1972683}
    &  &
    &  &  &  &  &      & \cellcolor{black} & \rotcheckmark & \rotcheckmark & \rotcheckmark &
    &
        & \rotcheckmark & \rotC &  & \rotH
\\
\rowcolor{black!10} \cellcolor{white}
\cite{Camera:2005:IDE:1085130.1085145} \bf{BORPH}
    & \rotcheckmark & \rotcheckmark
    & \rotcheckmark &  & \rotcheckmark & \rotcheckmark     & \rotcheckmark
    & \rotcheckmark &  &  &  &
    &
        & & \rotC &  & \rotH
\\
\midrule
(see~\S\ref{phd:design})
\bf{PhD} & \rotcheckmark & \rotcheckmark
    & \rotcheckmark & \rotcheckmark & \rotcheckmark &
    & \rotcheckmark
    & \rotcheckmark & \rotcheckmark  &  &  & \rotcheckmark
    & \rotcheckmark
        & \cellcolor{black} & \rotC &  & \cellcolor{black}

\end{tabular}
\end{center}
\caption{Survey of features provided by debugging systems. Blacked-out boxes
mean ``not applicable''.}
  \label{tab:survey}
  \vspace{-1em}
\end{table*}

Table~\ref{tab:survey} surveys closely related work, classifying it according to
the features they provide. Related work is contrasted in more detail
in~\S\ref{sec:relwork}.
In Table~\ref{tab:survey} \emph{extension points} means being able to extend the program
at certain points at runtime. \emph{Interruption} refers to asynchronous
interruption of a program by a debugger. \emph{Fine granularity} means that
we can look at arbitrary parts of a program; for instance
\citet{McKechnie:2009:DFP:1543136.1542470} only allow inspecting at the module
boundary.
\emph{Software instance} means that the program can be run on the CPU
as a process, independent from the FPGA.
\emph{Network/Control} refers to how a technique is implemented: as a control
loop or as a network that feeds signals to a logic analyser.
\emph{Use leftover resource} means that the debug circuitry does not compete
with the program's circuit. And \emph{embed at Source/HDL} refers to whether
the source code or its HDL image is updated to include the debug circuitry.

\subsection{PhD design features}
\label{phd:design}
PhD came about after we found ourselves extending our \emph{ad hoc} debugging
and monitoring code to support additional features, instigating us to study
the problem more rigorously.

Table~\ref{tab:survey} shows the features supported by PhD.  Initially it
supported state inspection and update. Updating state enabled us to influence
data-dependant control flow. Extending the controller to include branching
and extension points enabled us to support more features, such as assertion
checking, which involves breaking if a condition is satisfied.

As an idea, PhD is neither committed to HLS nor HDL description. In our
prototype we implemented it in HLS for convenience, but often better performance
can be gained by relying partly or fully on modules written in HDL.

\section{A language and model for program directing}
\label{sec:emu:directing}
In this section we describe $\mathfrak{D}$, a language of direction commands that
we extracted by analysing the commands commonly given to profilers
and debuggers such as gdb.\footnote{https://www.gnu.org/software/gdb/} In
Figure~\ref{example:tracing:1} we described one such command and how we code the high-level direction command as a program that will be
executed by the controller. 

\subsection{Direction language $\mathfrak{D}$}
\label{lang:direct}

\newcommand{\isconditional}{\ensuremath{\langle B \rangle}\xspace}
\newcommand{\isresourceful}{\ensuremath{\langle \$ \rangle}\xspace}

\newcommand{\sP}{\ensuremath{\;}}
\newcommand{\dirPrint}[1]{\ensuremath{\mathsf{print} \sP #1}}
\newcommand{\dirBreak}[3]{\ensuremath{\mathsf{break} \sP #1 \sP #2 \sP #3}}
\newcommand{\dirUnbreak}[1]{\ensuremath{\mathsf{unbreak} \sP #1}}
\newcommand{\dirStep}{\ensuremath{\mathsf{step}}} \newcommand{\dirSnapshot}{\ensuremath{\mathsf{snapshot}}} \newcommand{\dirBacktrace}[1]{\ensuremath{\mathsf{backtrace} \sP #1}}
\newcommand{\dirWatch}[2]{\ensuremath{\mathsf{watch} \sP #1 \sP #2}}
\newcommand{\dirUnwatch}[1]{\ensuremath{\mathsf{unwatch} \sP #1}}

\newcommand{\dirWhen}{\ensuremath{\mathsf{when}}}
\newcommand{\dirFor}{\ensuremath{\text{\textsf{for}}}}

\newcommand{\dirBreakVerbose}[3]{\ensuremath{\mathsf{break} \sP #1 \sP \dirWhen
\sP
#2 \sP \dirFor \sP #3}}
\newcommand{\dirWatchVerbose}[2]{\ensuremath{\mathsf{watch} \sP #1 \sP \dirWhen
\sP #2}}

Our direction language $\mathfrak{D}$ consists of the commands listed in
Table~\ref{tab:direction-commands}. These commands can have three kinds of
parameters: (i)~symbols relating to the program, (such as variable or function names,
or labels), (ii)~relations over program variables, which we denote by the
symbol \isconditional, indicating the conditions on which
statements might apply, and (iii)~measures of resource, which we symbolise by
\isresourceful, indicating a finite resource that is allocated for the execution
of a command.
Symbol $X$ is a metavariable ranging over variable identifiers in the
source program, and $L$ is a metavariable ranging over
\emph{labels} in the source program. A label is associated with a single
position in the program, (e.g., line 5 in function ``main''), but a single position
might be associated with multiple labels. below.

Let $\mathfrak{D}\isconditional$ be the set of possible conditions that can be
used. We assume that at least $\mathsf{true} \in \mathfrak{D}\isconditional$.
We can also allow additional truth conditions, and in this model we will have $\left(V_1=V_2\right) \in
\mathfrak{D}\isconditional$ for arbitrary $V_1,V_2$ ranging over program values
or variables. For example, ``$\dirWatch{v}{(v=5)}$'' would instruct the controller
to watch a variable $v$, and switch to interactive directing when $v=5$.

$\mathfrak{D}\isresourceful \in \mathbb{N}$ describes the maximum
quantity of some resource when carrying out a direction command.
This value must be less than the compile-time allocation of the resource, to ensure
the provision of sufficient resource for the command at runtime. This is needed to size the buffers used for tracing.
For example,
``$\mathsf{count} \sP \mathsf{reads} \sP v \sP \mathsf{true} \sP 5000$''
will count the number of reads of $v$, and break after 5000 reads have been
made. This could be done to avoid overflow, or to capture some behaviour of
interest.
Similarly,
``$\mathsf{trace} \sP \mathsf{start} \sP v \sP \mathsf{true} \sP 500$''
breaks after 500 instances of $v$ have been recorded in the trace. The trace
buffer must accommodate at least 500 entries.

\newcommand{\dirPos}{\ensuremath{\mathit{pos}}}

\newcommand{\vdecl}{\ensuremath{\mathit{vdecl}}}
\newcommand{\fdecl}{\ensuremath{\mathit{fdecl}}}
\newcommand{\return}[1]{\ensuremath{\mathrm{return}\;#1}}
\newcommand{\fname}{\ensuremath{\mathit{fname}}}
\newcommand{\ifthen}[2]{\ensuremath{\mathrm{if} \; #1 \; \mathrm{then} \; #2}}
\newcommand{\binop}[2]{\ensuremath{#1 \; \mathit{op} \; #2}}
\newcommand{\skipCommand}{\ensuremath{\mathrm{skip}}}
\newcommand{\extend}[1]{\ensuremath{\mathrm{extend}\{#1\}}}

\begin{table*}
\begin{tabular}{ll|l}
\toprule
  \multicolumn{2}{c}{\bf{Command}} & \bf{Behaviour} \hspace{11cm} \\
\midrule
  \multicolumn{2}{l|}{\dirPrint{X}} & Print the value of variable $X$ from the
  source program.\\
  \multicolumn{2}{l|}{\dirBreak{L}{\isconditional}{}} & Activate a
  (conditional) breakpoint at the position of label $L$. \\
  \multicolumn{2}{l|}{\dirUnbreak{L}} & Deactivate a breakpoint.\\
  \multicolumn{2}{l|}{\dirBacktrace{\isresourceful}} & Print the ``function call stack''.\\
  \multicolumn{2}{l|}{\dirWatch{X}{\isconditional}} & Break when $X$ is updated
  and satisfies a given condition.\\
  \multicolumn{2}{l|}{\dirUnwatch{X}} & Cancel the effect of the ``watch''
  command.\\
  $\mathsf{count}$ & $\hspace{-3mm}\left\{
  \begin{array}{l}
    \mathsf{reads} \sP X \sP \isconditional \sP \isresourceful\\
    \mathsf{writes} \sP X \sP \isconditional \sP \isresourceful\\
    \mathsf{calls} \sP \fname \sP \isconditional \sP \isresourceful\\
  \end{array}\right\}$ & Count the reads or writes to a variable $X$, or the calls
to a function $\fname$. \\
  $\mathsf{trace}$ & $\hspace{-3mm}\left\{
  \begin{array}{l}
  \mathsf{start} \sP X \sP \isconditional \sP \isresourceful\\
  \mathsf{stop} \sP X\\
  \mathsf{clear} \sP X\\
  \mathsf{print} \sP X\\
  \mathsf{full} \sP X\\
  \end{array}\right.$ &
  \begin{tabular}{l}
    Trace a variable, subject to a condition being satisfied, and up to trace some length.\\
    Stop tracing a variable.\\
    Clear a variable's trace buffer.\\
    Print the contents of a variable's trace buffer.\\
    Check if a variable's trace buffer is full.\\
  \end{tabular}
\end{tabular}
  \caption{\label{tab:direction-commands}
  Directing commands making up language $\mathfrak{D}$.
  Note that \textsf{count} has similar subcommands to those of \textsf{trace},
  to clear the counters, get their current value, and find out if a maximum value
  has been reached.}
\vspace{-1em}
\end{table*}

\subsection{Controller}
\label{casp-machines}
High-level direction commands such as those in Table~\ref{tab:direction-commands}
are ultimately translated into programs that run on a simple controller
embedded in the program.
We model the controller as a \emph{CASP} machine (for ``Counters, Arrays,
and Stored Procedures'', the constituents of the machine's memory). 
CASP machines are very weak. They are more structurally complex
than register machines~\cite{Shepherdson:1963:CRF:321160.321170} since they
have separate memories for storing arrays and registers, but CASP machines are
computationally much weaker than register machines, unable to encode
partial computable functions.
A limited form of memory indirection is permitted through a collection of arrays.
The language lacks any means for defining recursive functions, or branching to arbitrary addresses.
Any more complex computation must be done by the director; the controller simply
provides a controlled access to the program's memory.

We describe the language of CASP programs in Figure~\ref{fig:casp-syntax}.
We rely on the following meta-variables and syntax categories:
$P$ programs,
$E$ expressions,
$I$ indices,
$U$ updatable values,
$V$ values,
$N$ numerals (corresponding to $\mathbb{Z}$),
$X$ variable identifiers, and
$R$ the array identifiers, where the names for variables and those for arrays
are disjoint, $X \cap R = \emptyset$.

\newcommand{\CaspBreak}{\ensuremath{\mathtt{break}}\xspace}
\newcommand{\CaspCont}{\ensuremath{\mathtt{continue}}\xspace}
\begin{figure}
\begin{minipage}[t]{0.25\textwidth}
\[
\begin{array}{rcll}
  P & ::= & E \\
    &  |  & U := E\\
    &  |  & \mathit{op} \; U \\
    &     & \mathit{op} \in \{\mathtt{inc}, \mathtt{dec}\} \\
    &  |  & P_1; P_2 \\
    &  |  & \mathtt{if} \; E \; \mathtt{then} \; P_1 \\
    &     & \hspace{1.5mm} \qquad \mathtt{else} \; P_2 \\
    &  |  & \CaspBreak \\
    &  |  & \CaspCont \\
    &  |  & @L:\{P'\}
\end{array}
\]
\end{minipage}
\hspace{-5mm}\begin{minipage}[t]{0.25\textwidth}
\[
\begin{array}{rcll}
  E & ::= & V \\
    &  |  & -V\\
    &  |  & V_1 \; \mathit{op} \; V_2\\
    &     & \mathit{op} \in \{=, <\} \\
  I & ::= & N \\
    &  |  & X\\
  U & ::= & X \\
    &  |  & R[I]\\
  V & ::= & I \\
    &  |  & R[I]
\end{array}
\]
\end{minipage}
  \caption{\label{fig:casp-syntax} Syntax for CASP programs.}
  \vspace{-1em}
\end{figure}

$@L:\{P\}$ is a \emph{placement} command: it updates the code at extension point
having label $L$ to be $P$. Note that placement commands may not be nested in our model: for
instance, this is not a valid program:
\vspace{-0.5em}
\[\begin{array}{l}
  @L: \left\{\mathtt{if} \; x = 1 \; \mathtt{then} \; @M:
\left\{\mathtt{break}\right\} \right. \\
\left. \qquad\qquad\qquad \;\; \; \mathtt{else} \; @M:
\left\{\mathtt{continue}\right\}\right\}
\end{array}\]
We do not want to allow programs to be self-modifying in this way, since it
complicates reasoning about them.

\paragraph{Ending programs.}Both \CaspCont and \CaspBreak indicate the end of a CASP program, but differ
in what happens before resuming the host program (in which the controller is
embedded). \CaspCont simply resumes
where the host program left off, whereas  \CaspBreak switches into an
interactive direction mode. In this mode, the controller may receive commands
from the director, execute them, and send an acknowledgement back.

The remaining syntax forms and constants used above are standard and intuitive.
Owing to its simplicity the semantics of this language are straightforward,
and are given in~\S\ref{casp-machines:semantics}.

\subsection{Examples}
\label{sec:casp:examples}
CASP programs are to $\mathfrak{D}$ what microprograms are to an
Instruction Set Architecture~\cite{Smith:1974:MRA:1008311.1008315}.
We give some examples of coding program direction commands below,
before describing the behaviour of program direction commands in more detail
in the next two sections.

\paragraph{Conditional tracing.}
Let \texttt{BASIC} represent the program that codes the behaviour of
``$\mathsf{trace} \; V ...$'' from Figure~\ref{example:tracing:1}.
We can code the \emph{conditional} variant of this command, for example when
$V$ is less than some threshold value $\mathrm{V\_trace\_threshold}$ (and where
$>=$ is syntactic sugaring):
\begin{lstlisting}[belowskip=-0.5\baselineskip, basicstyle=\small\ttfamily,
    escapeinside={@}{@}]
if V >= V_trace_threshold then
  continue
else
  BASIC
\end{lstlisting}

\paragraph{Sampled tracing.}
The \emph{sampled} variant involves allocating an additional variable to count
the interval between samples, and storing the desired sample interval:
\begin{lstlisting}[belowskip=-0.5\baselineskip, basicstyle=\small\ttfamily,
    escapeinside={@}{@}]
if V_trace_samp = 0 then
  V_trace_samp := V_trace_samp_interval;
  BASIC
else
  dec V_trace_samp;
  continue
\end{lstlisting}

\paragraph{Profiling.}
The command ``$\mathsf{count} \; \mathsf{writes} \; v$'' causes a counter
to be incremented each time a variable is updated:
\begin{lstlisting}[belowskip=-0.5\baselineskip, basicstyle=\small\ttfamily,
    escapeinside={@}{@}]
if V_count_writes > V_count_writes_max then break
else
  inc V_count_writes;
  continue
\end{lstlisting}

\paragraph{Watchpoints.}
The command ``$\mathsf{watch} \; v \; \isconditional$'' causes the program
to break (for interactive guidance) when a variable's value satisfies some
predicate.
Let $B$ be the CASP-level value to the $\mathfrak{D}$-level \isconditional
parameter (we give such a function in~\S\ref{theory-rules:break}).
The code is similar to that in the profiling example above, except
that $B$ is user-provided.
\begin{lstlisting}[basicstyle=\small\ttfamily,
    escapeinside={@}{@}]
if @$B$@ then break else continue
\end{lstlisting}

\paragraph{Breakpoints.}
``$\mathsf{break} \; L \; \isconditional$'' causes the program to break
when it reaches a specific label, and if condition $B$ is satisfied.
The coding is identical to that in the watchpoint example, but they differ in
their \emph{placement}: breakpoints are placed at programmer-specified positions
in the code whereas watchpoints are associated with labels where variables
are updated.  This difference cannot be seen from the snippet, but will become
evident in the formalisation of the program direction commands, which we start
next.

\subsection{Directability ordering}
\label{directability-ordering}
\newcommand{\DState}[1]{\ensuremath{\mathcal{#1}}}
\newcommand{\CState}[1]{\ensuremath{\mathcal{#1}}}
\newcommand{\Prog}[1]{\ensuremath{#1}}
In this section we define a relation $x \sqsubset x'$ to mean
``$x'$ is more directable than $x$'', where $x, x'$ are triples
$\left(\DState{D}, \CState{C}, \Prog{p}\right)$ and
$\left(\DState{D}', \CState{C}', \Prog{p}'\right)$,
each representing three interdependent parties: the \emph{director},
\emph{controller} and \emph{program}. The user (or their agent) issues direction
commands to the director, which interacts with the program's state via its
agent, the controller, embedded in the program.
We use this relation to give semantics to the direction commands in terms of
interaction with CASP machines.

Our directability relation gathers information about the three parties involved,
and describes how their interdependence is revealed by the directing commands:
for example, the director would not be able to execute $\mathsf{trace} \; X$
if the program did not have a variable called $X$, or if the controller had not
been allocated a trace buffer.

In our notation, $\DState{D}$ represents the director's state (a set of facts
representing its knowledge about the controller's state, such as which
breakpoints exist, and whether they are active or not). $\CState{C}$ is the
controller's state, consisting of a $(C,A,\mathit{SP})$ machine, cf
\S\ref{casp-machines}. $\Prog{p}$ is a program. We will define an example
language in \S\ref{lang:prog} to aid our formalisation.

We also include in the relation some information about \emph{why}
one triple is less directable than the other. We therefore index the relation
by (i) $\mathfrak{C} \subseteq \mathfrak{D}$ the direction commands (\S\ref{lang:direct}) supported by
$\left(\DState{D}, \CState{C}, \Prog{p}\right)$, (ii) $c \in \mathfrak{D}$ the additional command
supported by $\left(\DState{D'}, \DState{C'}, \Prog{p'}\right)$, and $D_c \in
\left(\DState{D'} \to \DState{D'}\right)$ the semantics of this command.
Note that $\mathfrak{C}$ denotes the set of direction commands that is supported
\emph{simultaneously} by $\left(\DState{D}, \CState{C}, \Prog{p}\right)$, i.e.,
these commands are allocated separate state.

Written out in full, we obtain this relation:
\newcommand{\directorder}[2]{\ensuremath{\stackrel{#2}{\sqsubset_{#1}}}}
\[
  \left(\DState{D}, \DState{C}, \Prog{p}\right) \;
  \directorder{\mathfrak{C}}{c} \;
  \left(\DState{D'}, \DState{C'}, \Prog{p'}\right) \; : \; D_c
\]

In~\S\ref{theory-rules} we will instantiate such a relation by formalising
commands from $\mathfrak{D}$ in terms of CASP machines. Note that this describes
how the directing commands are translated into CASP programs, but we do not
fully formalise the director: $D_c$ is written in an ML-like pseudocode.

Our formalisation is devised in away that avoids the mutual interference of
direction commands. That is, the same program can be subjected to any
combination of direction commands.
\newcommand{\DStateDiff}[1]{\breve{\DState{#1}}}
\newcommand{\CStateDiff}[1]{\breve{\CState{#1}}}
To make this non-interference more precise, we introduce some definitions.
Let $\DStateDiff{D} = \DState{D'}\backslash \DState{D}$ and $\CStateDiff{C} =
\CState{C'}\backslash \CState{C}$.
We say that $D_c$ is \emph{relevant} to $\DState{D'}\backslash\DState{D}$
if it only manipulates state or elements introduced in $\DState{D'}$ and
$\CState{C'}$.
Furthermore commands are \emph{disjoint}
if they introduce non-overlapping state. That is, for any two commands
$c_1$ and $c_2$, for any prior states $\CState{C}$ and $\DState{D}$ their
respective new states are disjoint:
$(\DState{D}_1 \backslash \DState{D}) \cap (\DState{D}_2 \backslash \DState{D})
= \emptyset =
(\CState{C}_1 \backslash \CState{C}) \cap (\CState{C}_2 \backslash \CState{C})$.

\subsubsection{Program language}
\label{lang:prog}
In this section we specify a first-order imperative language to support our
formalisation of program direction commands.
Unlike CASP machines~(\S\ref{casp-machines}) this language is computationally
strong: recursive functions over the integers can be encoded.
The language's simplicity enables the relation of program direction with CASP
machines, while avoiding excessive formal complexity.
Formalising transformations for realistic languages---even simple
transformations~\cite{Schafer:2009:CPV:1481848.1481859}---is usually fraught with
complex definitions, and we avoid that here.

The language grammar is given next.
Note that for simplicity we deliberately overlay the meta-variables for
variables and numerals ($X$ and $N$) over those for CASP machines. This
simplifies the interface with CASP programs, which will be executed at extension
points within host programs. $p$ ranges over programs, $s$ over statements, $e$
over expressions, $\tau$ over types, and $\vdecl$ and $\fdecl$ over the
declaration of variables and functions respectively.
\vspace{0.5em}
\[
\begin{array}{rcll}
  p & ::= & \overline{\vdecl} \; \overline{\fdecl} \; \return{\fname(\bar{e})} \\
  \vdecl & ::= & \tau \; X\\
  \fdecl & ::= & \tau \; \fname(\bar{x})\{\overline{s}; \return{e}\}\\
  s & ::= & \mathrm{skip} \\
    &  |  & X := e\\
    &  |  & \mathrm{if} \; e \; \mathrm{then} \; s;\overline{s} \\
    &  |  & \mathrm{extend}\{L_1,\ldots,L_n\} \\
  e & ::= & N \\
    &  |  & X\\
    &  |  & \fname(\bar{e})\\
    &  |  & e_1 \; \mathit{op} \; e_2\\
    &     & \mathit{op} \in \{+, -, ==, <\} \\
\end{array}
\]

In this language the only data type in $\tau$ is the integer type.
The only unusual construct in this language is
$\mathrm{extend}\{L_1,\ldots,L_n\}$.
This indicates an \emph{extension point}, where control is passed to
the CASP machine~(\S\ref{casp-machines}). As before, $L$ is a metavariable
ranging over \emph{labels}
drawn from a denumerable set.

The semantics of this language are straightforward, and are given
in~\S\ref{lang:prog:semantics}.
Intuitively, the behaviour of $\mathrm{extend}\{L_1,\\\ldots,L_n\}$ is as follows.
If $n=0$ then the command has no effect. Otherwise, the stored procedure
associated with each $L_i$ is called, in any order, and run to completion,
noting the last instruction of each $L_i$. The last instruction of any CASP
program is either `\texttt{continue}' or `\texttt{break}'~(\S\ref{casp-machines}).
If all $L_i$ end in `\texttt{continue}', then the behaviour of
$\mathrm{extend}\{L_1,\ldots,L_n\}$ is to continue executing the next statement
in the host program.
Otherwise, if at least one $L_i$ ends in `\texttt{break}', then the controller
switches into interactive mode. In this mode, control remains with the
controller, until the director sends it a `\texttt{continue}' command, at which
point control is returned to the program.

We make the simplifying assumption that all the statements in the user's program
are interspersed with `$\mathrm{extend}$': that is, if the user writes
$s_0;\ldots;s_n$
then this is translated into
$\mathrm{extend}\{\};s_0;\mathrm{extend}\{\};\ldots;\mathrm{extend}\{\};s_n;\mathrm{extend}\{\}$.
In the next section we populate these extension points with labels, to extend
the directability of a program.

\subsection{Semantics for $\mathfrak{D}$}
\label{theory-rules}
In this section we encode program direction commands~(\S\ref{lang:direct}) into
interactions between the director and controller~(\S\ref{casp-machines}).
Note that this describes how the directing commands are translated into CASP
programs, but we do not fully formalise the director: $D_c$ is written in an
ML-like pseudocode.

\newcommand{\directability}[4]{  #1 \;
  \directorder{#3}{#4} \;
  #2 \;
}
\newcommand{\directTwo}[3]{\left\{
\begin{array}{l}
  \DStateDiff{D} = \{ #1 \}\\
  \CStateDiff{C} = \left\{ #2 \right\}\\
  D_c = \lambda \DState{D'}.\raisebox{3mm}{\adjustbox{valign=T}{$#3$}}
\end{array}\right.
}

\subsubsection{Break}
\label{theory-rules:break}
We start by formalising the meaning of the ``break'' command. Intuitively, this
command adds a breakpoint to a program: an extension point is at the position of
the breakpoint is labelled with $L$, and the associated state is set up in the
director and controller.
To support this command:
\begin{itemize}
  \item A program $p$ is extended to include a label $L$ at the position where
    the breakpoint is to be placed. This extension is formalised by the premise
    $p <^1_L p'$, which means that $p$ is identical to $p'$ except for the label
    $L$ occurring at some extension point. This is defined formally in~\S\ref{adx:metaprog} along with
    related definitions, such as that of a \emph{position} in the program.

  \item The label $L$ must not appear in the original program. We write this as
    $L \not\in p$ using an abbreviation defined in~\S\ref{adx:metaprog}.

  \item The controller's state is extended to store the procedure associated
    with $L$. Furthermore, the breakpoint is activated by default.
    We use the abbreviated notation
    $\CStateDiff{C} = \{\mathit{SP}[L\mapsto\mathtt{break}]\}$ to indicate this
    extension, where $\mathit{SP}$ is the stored-procedure memory in the CASP.

  \item The director's \emph{state} is extended to encode whether the breakpoint is
    currently active or not. It is activated by default, thus:
    $\DStateDiff{D} = \{(\text{<<bp>>},L,1)\}$, where ``<<bp>>'' is a unique
    token we use for breakpoints, and $1$ is a token we use to indicate that the
    breakpoint is active. Soon we will formalise the ``unbreak'' command, which
    flips this value to $0$.

  \item The director's \emph{behaviour} for the ``break'' command, $D_c$,
    involves activating the breakpoint unless already active.
\end{itemize}

\newcommand{\dirBreakTwo}[2]{\ensuremath{\mathsf{break} \sP #1 \sP #2}}

\newcommand{\fatbrackSP}[1]{\fatbrack{#1}_{\mathit{SP}}}

In the definition of $D_c$ below we use the notation $P \leadsto N$, which we use to
mean that the director sent the CASP program $P$ to the controller, and received
the reply $N$. Thus ``$@L:\{\mathtt{break}\}\leadsto\ulcorner L\urcorner$''
means that the director instructed the controller to store the program
$\mathtt{break}$ at $L$, and that it expects to get $\ulcorner L\urcorner$ back
(which is a code indicating where the program is stored, as formalised
in~\S\ref{casp-machines:semantics}).

In $D_c$ we use the notation
``$(\text{<<bp>>},L,0 \mapsto 1):\in\mathcal{D'}$'' to abbreviate
  $\{(\text{<<bp>>},L,1)\}\cup\left(\mathcal{D'}\backslash\{(\text{<<bp>>},L,0)\}\right)$.

The formalisation of the ``break'' command follows.
We can use a simplified notation since our commands will be both relevant to
$\DState{D'}\backslash \DState{D}$ and disjoint:
\begin{mathpar}
\resizebox{.45 \textwidth}{!}{
  \inferrule
    {L \not\in p \\ p <^1_L p'}
    {\directability{p}{p'}{\mathfrak{C}}{\dirBreakTwo{L}{\isconditional}}
    \directTwo{(\text{<<bp>>},L,1)}{\mathit{SP}[L\mapsto \fatbrackSP{\dirBreakTwo{L}{\isconditional}} ]}
    {\begin{array}{l}
      \mathrm{if} \; (\text{<<bp>>},L,1) \in \mathcal{D'} \; \mathrm{then} \;
    \mathcal{D'}\\
      \mathrm{else} \\
      \quad \fatbrack{\dirBreakTwo{L}{\isconditional}} \leadsto\ulcorner
    L\urcorner;\\ \quad
       (\text{<<bp>>},L,0 \mapsto 1):\in\mathcal{D'}
    \end{array}}}}
\end{mathpar}
where
\[\begin{array}{l}
\fatbrackSP{\dirBreakTwo{L}{\isconditional}}
= \mathsf{conditional} \; \isconditional \;\mathtt{break} \\
\fatbrack{\dirBreakTwo{L}{\isconditional}}
= @L:\{\fatbrackSP{\dirBreakTwo{L}{\isconditional}}\} \\
\end{array}\]

We use $\fatbrack{\dirBreakTwo{L}{\isconditional}}$ to denote the meaning of \\
``\dirBreakTwo{L}{\isconditional}'' to the director, as a CASP program.\\
$\fatbrackSP{\dirBreakTwo{L}{\isconditional}}$ is the value used to initialise
the stored program associated with $L$.

The meaning of \isconditional is translated for inclusion in the CASP program by
the following function:
\[
  \begin{array}{l}
  \mathsf{conditional} \; \isconditional \; t \; = \\
    \quad
  \begin{cases}
    t & \text{if } \isconditional = \mathsf{true} \\
    \kern-2mm    \begin{array}{l}
    \mathtt{if} \; I_1 == I_2 \; \mathtt{then} \; t \\
      \;\;\quad \mathtt{else} \; \mathtt{continue}
    \end{array} & \raisebox{3mm}{$\text{if } \isconditional = \left(I_1 =
    I_2\right)$} \\
  \end{cases}
  \end{array}
\]

We now turn to the ``unbreak'' direction.
To be able to issue this direction, the breakpoint needs to exist---thus we have
a dependency on the ``break'' direction earlier:
\begin{mathpar}
\resizebox{.45 \textwidth}{!}{
  \inferrule
    {\left(\dirBreakTwo{L}{\isconditional}\right) \in \mathfrak{C}}
    {\directability{p}{p}{\mathfrak{C}}{\dirUnbreak{L}}
    \directTwo{}{}
    {\begin{array}{l}
      \mathrm{if} \; (\text{<<bp>>},L,0) \in \mathcal{D'} \; \mathrm{then} \;
    \mathcal{D'}\\
      \mathrm{else} \; \fatbrack{\dirUnbreak{L}}\leadsto\ulcorner
    L\urcorner;\\ \quad
    (\text{<<bp>>},L,1 \mapsto 0):\in\mathcal{D'}
    \end{array}}}}
\end{mathpar}
where
\[\begin{array}{l}
\fatbrack{\dirUnbreak{L}}
  = @L:\{\mathtt{continue}\} \\
\end{array}\]
Note that the program, and the states of the controller and director are not
changed. The only extension is made to the behaviour of the director, which is
extended with a function to unset the breakpoint.
This time we didn't set anything in the controller since there is no default
behaviour and additional state required for unsetting a breakpoint. This is
because unbreaking a breakpoint doesn't establish the breakpoint,
whereas breakpoint does.

To print the value of a variable $X$, that variable needs to exist in
the program ($X \in \mathrm{Var}_p$), and we need to have at least one extension
point (through which we can send the print command).
\begin{mathpar}
  \inferrule
    {\left(\dirBreakTwo{L}{\isconditional}\right) \in \mathfrak{C} \\
      X \in \mathrm{Var}_p
     }
    {\directability{p}{p}{\mathfrak{C}}{\dirPrint{X}}
    \directTwo{}{}
    {\begin{array}{l}
      X \leadsto N; \\
      \mathrm{print}(N)
    \end{array}}}
\end{mathpar}
This time we didn't use a placement command to update the behaviour of the
controller; we simply ran a query.
$\mathrm{print}(N)$ is pseudocode that uses a print function in the director.
Recall that we formalise directions in terms of CASP machines, and don't formalise the
director's behaviour.
$X \in \mathrm{Var}_p$ ensures that $X$ is a variable in $p$. $\mathrm{Var}_p$ is defined in~\S\ref{adx:metaprog}.

\newcommand{\dirTraceStartThree}[3]{\ensuremath{\mathsf{trace} \sP \mathsf{start}
\sP #1 \sP #2 \sP #3}}
\newcommand{\dirTraceStop}[1]{\ensuremath{\mathsf{trace} \sP \mathsf{stop} \sP #1}}
\newcommand{\dirTracePrint}[1]{\ensuremath{\mathsf{trace} \sP \mathsf{print} \sP #1}}
\newcommand{\dirTraceFull}[1]{\ensuremath{\mathsf{trace} \sP \mathsf{full} \sP #1}}
\newcommand{\dirTraceClear}[1]{\ensuremath{\mathsf{trace} \sP \mathsf{clear} \sP #1}}

\newcommand{\wideDirectability}[7]
{\begin{array}{c}
    \directability{#1}{#2}{#3}{#4} \\
    \directTwo{#5}{#6}
    {#7}
\end{array}}

\subsubsection{Trace}
\label{theory-rules:trace}
The most important command related to tracing is ``trace start''; the other
commands depend on it.
\begin{mathpar}
\resizebox{.5 \textwidth}{!}{
  \inferrule
    {X \in \mathrm{Var}_p \\ \forall L \in X_L. \; L \not\in p \\
    \mathrm{Positions}_{p'}(X_L) = \mathrm{PostUpdate}_{p'}(X) \\ p <_{X_L} p'}
    {\directability{p}{p'}{\mathfrak{C}}{\dirTraceStartThree{X}{\isconditional}{\isresourceful}}
}}
\end{mathpar}
\[      \directTwo{(\text{<<t>>},X,1)}{
    \begin{array}{l}\mathit{C}[X_\mathrm{i}]=0, \mathit{C}[X_\mathrm{of}]=0,
      \mathit{A}[X_\mathrm{a}[\isresourceful]],\\
      \mathit{SP}[L \mapsto \fatbrackSP{\dirTraceStartThree{X}{\isconditional}{\isresourceful}}]\\
      \quad \text{for each $L \in X_L$}
      \end{array}}
    {\begin{array}{l}
      \mathrm{if} \; (\text{<<t>>},X,1) \in \mathcal{D'} \; \mathrm{then} \;
    \mathcal{D'}\\
      \mathrm{else} \\
      \quad \text{for each $L \in X_L$:}\\
      \qquad @L:\{\fatbrackSP{\dirTraceStartThree{X}{\isconditional}{\isresourceful}}\}\\
      \qquad\qquad \leadsto\ulcorner L\urcorner;\\
      \quad (\text{<<t>>},X,0 \mapsto 1) :\in\mathcal{D'}
    \end{array}}
    \]
    \[
      \begin{array}{l}
\fatbrackSP{\dirTraceStartThree{X}{\isconditional}{\isresourceful}} = \\
\qquad\qquad    \quad \mathsf{conditional} \; \isconditional \\
\qquad\qquad    \qquad \left(
    \begin{array}{l}
      \mathtt{if} \; X_\mathrm{i} < \isresourceful \; \mathtt{then} \\
        \qquad X_\mathrm{a}[X_\mathrm{i}] := X;\\
        \qquad \mathtt{inc} \; X_\mathrm{i}; \\
        \qquad \mathtt{continue} \\
      \mathtt{else} \\
        \qquad \mathtt{inc} \; X_\mathrm{of}; \\
        \qquad \mathtt{break}
    \end{array} \right)
\end{array}
\]

Applying this rule depends on a set of labels $X_L$ in $p'$ that don't exist
in $p$.  $p'$ is the least extension of $p$ that includes these labels.
Furthermore, these labels coincide with the positions in the program
occurring after $X$ has been updated.

The controller's state is extended with the buffer index,
$X_\mathrm{i}$, initialised to 0; the overflow indicator
$X_\mathrm{of}$, initialised to 0 (for false); and
$X_\mathrm{a}$ is an array that can hold $\isresourceful$ elements.
Each $L \in X_L$ labels the positions immediately after the variable has been
updated.

\begin{mathpar}
  \inferrule
    {\left(\dirTraceStartThree{X}{\isconditional}{\isresourceful}\right) \in \mathfrak{C}
    \\ \mathrm{Positions}_{p}(X_L) = \mathrm{PostUpdate}_{p}(X)}
    {\directability{p}{p}{\mathfrak{C}}{\dirTraceStop{X}}}
\end{mathpar}
\[
    \directTwo{}{}
    {\begin{array}{l}
      \mathrm{if} \; (\text{<<t>>},X,0) \in \mathcal{D'} \; \mathrm{then} \;
    \mathcal{D'}\\
      \mathrm{else} \\
      \quad \text{for each $L \in X_L$:}\\
      \qquad @L:\{\fatbrackSP{\dirTraceStop{X}}\} \leadsto\ulcorner L\urcorner;\\
    (\text{<<t>>},X,1 \mapsto 0):\in\mathcal{D'}
    \end{array}}
\]
where $\fatbrackSP{\dirTraceStop{X}}
  = \mathtt{continue}$.

The remaining commands are formalised in~\S\ref{adx:further-commands},
supported by program-level predicates and functions defined in~\S\ref{adx:metaprog}.
 
\section{Implementation and Evaluation}
\label{sec:eval}
We prototyped the ideas described in the previous section by extending network
programs with program-hosted directability, and compiled them to run on an FPGA.
We then evaluated the effect of these directability features on the program in
which they are embedded.
Our approach enables us to make fine-grained modifications to directability,
and we evaluate the overhead from supporting different CASP machine
instructions.

\subsection{Prototype}
In our prototype we manually transformed programs to include the controller.
This transformation was straightforward: we wrote the controller (implementing a
CASP machine) and added extension points to the program (consisting of calls to
the controller to execute a stored procedure).

\paragraph{Focus.}
From its description, the PhD idea is not constrained to a specific kind of program.
In our prototype we focussed on using it to work with \emph{network} programs however,
for two reasons:
\begin{enumerate}
\item It allows us to test remote directing over standard network
equipment.
In our survey of tools and techniques (Table~\ref{tab:survey}) only Dynascope
has network access, but it does not work for FPGAs.
\item It allows us to use industrial high-precision network
measuring equipment to see the effects on the program hosting a controller.
\end{enumerate}

\paragraph{Use-cases.}
As programs we used implementations of DNS and Memcached that we had written
previously to run on FPGAs, as part of earlier research. DNS (Domain Name
System) is a ubiquitous name-resolution system used on private and public
packet-switched networks such as the Internet~\cite{dns1}.
This implementation was around 700 lines of \csharp.
Memcached~\cite{memcached} is a well-known, in-memory key/value store that
caches read results in memory to quickly respond to queries. The protocol uses a
number of basic commands such as GET (retrieve a value associated with the
provided key), SET (store a key/value pair) and DELETE (remove a key/value pair)
and has both ASCII and binary protocols. In this work as proof-of-concept we have implemented a limited version of
Memcached supporting GET/SET/DELETE using the binary protocol over UDP and
supporting 6 byte keys and 8 byte values.
This implementation was almost 900 lines of \csharp.

\begin{figure}
  \centering
  \includegraphics[width=0.9\linewidth]{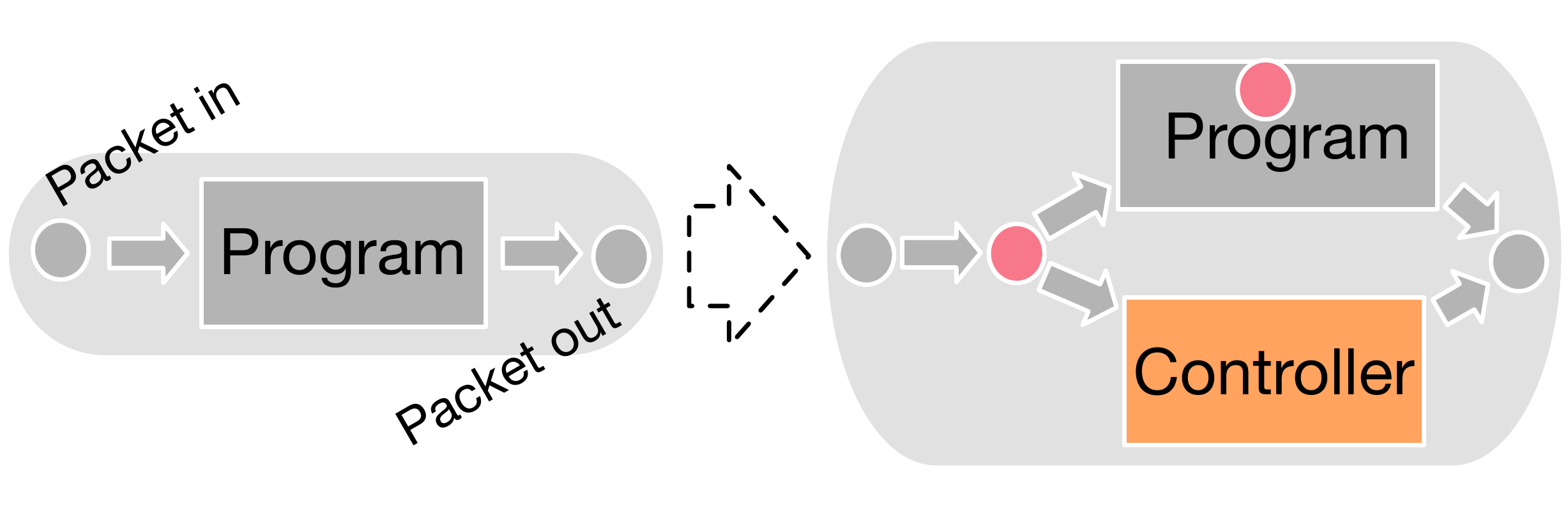}
  \caption{\label{fig:prototype}Transformation of the program to include a controller. Normal
  packets are handled as normal, but \emph{direction packets} are passed to the
  controller. Pink dots represent \emph{extension points}, one of which is added
  within the control flow of the original program in this illustration.}
\end{figure}

\paragraph{Method.}
We transformed the DNS and Memcached implementations in two ways:
(i)~adding code to check whether a received packet is a \emph{direction
packet} intended for the controller (see Figure~\ref{fig:prototype}), in which
case the controller (and not the original program) processes the packet;
(ii)~adding an extension point in the body of the (DNS or Memcached) main loop,
allowing us to influence and observe the program from that point.
We form an enumerated type that corresponds to the program variables whose
values the controller may access and change at runtime. The code for each value
of the enumerated type is used to refer to the program value, to instruct the
controller to increment it, for example. Building tool support to automate parts
of this process seems feasible.

\paragraph{Direction packets.}
Direction packets are network packets in a custom and simple packet format,
whose payload consists of (i)~code to be executed by the controller, or (ii)~status
replies from the controller to the director. It enables us
to remotely direct a running program, similar to gdb's `remote serial
protocol'.\footnote{\url{http://www.embecosm.com/appnotes/ean4/embecosm-howto-rsp-server-ean4-issue-2.html}}
Our design uses a simple direction language,  and works for instances of a
program that run both as software and hardware (on FPGAs), whereas gdb requires
special backends for each architecture.

\paragraph{Controller.}
Our controller follows the description of CASP machines very
closely~(\S\ref{casp-machines}). It has two features.
First, memory is organised into \emph{Counters}, \emph{Arrays}
    and \emph{Stored procedures}. Counters include variables in the original
    program, as well as extra registers used for program directing.
    In this prototype we only support numeric datatypes; structured datatypes
    could be encoded in principle.
Secondly, a function that interprets the language of CASP machines.
    This is used to branch to stored procedures when their corresponding
    extension points are reached.

\paragraph{Tools and equipment.}
We wrote our programs in \csharp, and used the Kiwi \emph{high-level synthesis}
(HLS) system~\cite{singh:08:kiwi} that statically recompiles .NET bytecode into
Verilog.
The Verilog code generated by Kiwi was slotted into open-source reference code from the
NetFPGA project,\footnote{\url{http://netfpga.org/}} and compiled to run on the
NetFPGA SUME board~\cite{zilberman2014netfpga}, a low-cost, \pcie host adapter
card able to support 40\Gbps and 100\Gbps applications. At the core of the board
is a Xilinx Virtex-7 690T FPGA device.
\begin{figure}
  \centering
  \includegraphics[width=0.9\linewidth]{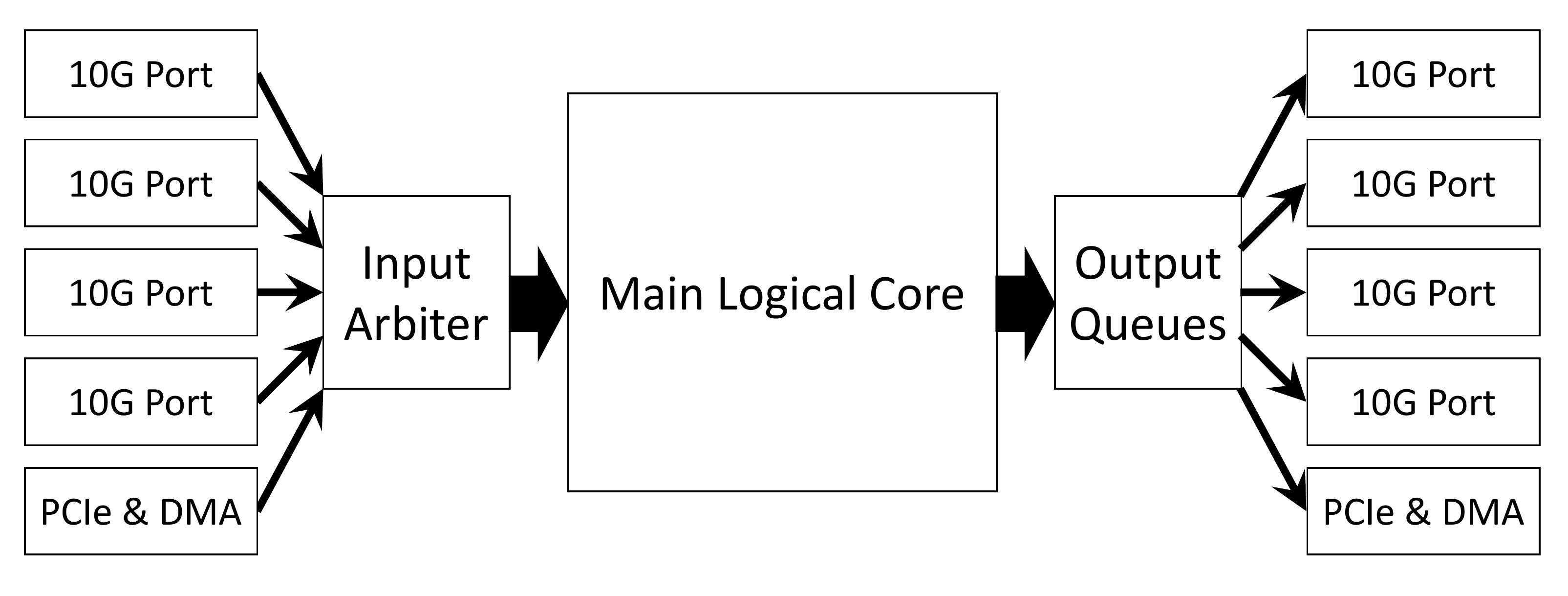}
  \caption{\label{fig:netfpga_pipeline}NetFPGA Reference Pipeline. Each of the
  implementations in our prototype consists of a separate Main Logical Core.}
  \vspace{-1em}
\end{figure}
The NetFPGA reference designs share the same FPGA architecture, illustrated in
Figure~\ref{fig:netfpga_pipeline}, of multiple physical interfaces surrounding a
logical data-path. Traffic to the FPGA enters through one of four 10\Gbps ports
or from a \pcie interface, and is passed into the main data-path by an input
arbiter, which arbitrates between all ingress interfaces. Packets are then
processed within the main logical core, and are diverted to their destination through an output queues
module. From this module, the packets are passed to the physical interfaces and
are transmitted.

\subsection{Evaluation}
We evaluate our prototype by carrying out a quantitative analysis of the impact
that the controller has on the program in which it is embedded. This impact is
measured in terms of \emph{utilisation} of resources on the FPGA, and the
\emph{performance} of the host program.

An FPGA consists of an interconnected grid of \emph{logic blocks}, which
in turn contain resources such as memory and logic functions (in the form of
so-called \emph{look-up tables}).
Flip-flops are primitive storage circuits.

Table~\ref{tab:comparison} shows the utilisation and performance for DNS and
Memcached, extended with different controller features. These features show
a fine-grained decomposition of the instructions supported by the controller:
reading a variable, writing a variable, and incrementing a variable. We see that
the impact on utilisation and performance is minimal.

Performance is analysed in three ways. The \emph{duration} is the number of
clock cycles that are needed by the program to process
a packet within the main logical core, as extracted from simulation. In our hardware, each clock cycle takes 10ns. \emph{Latency} is the time taken for
a program to service requests from the network. \emph{Throughput} is the
rate of requests that can be serviced by the program (before packets start to be
dropped).

Using the controller we can read and change the program's state at
the packet rate. Using JTAG and the Virtex-7 FPGA we can read data at up to 66Mbps, three orders of magnitude
less than the maximum throughput that the NetFPGA can sustain over high speed
serial interface --e.g. the PCIe channel. In principle the direction controller could
use any slice of that, subject to not interfering with the hosting program too much.

In Table~\ref{tab:more-comparison} we compare DNS extended with one extension
point, against DNS extended with an \emph{embedded logic analyser}~(ELA).
An ELA is a standard technique in hardware development, and consists of a
circuit that passively monitors the program, creating a trace of the program's
execution. DNS+2e is the DNS extended with one extension point that only
contains a NOP. In (Count) the extension point's stored procedure is changed
into a counter, and in (Trace) it is changed to emulate the behaviour of the
ELA.

These results confirm that the resources overhead is minimal,
making PhD a feasible solution. We note that in the use-cases detailed
below the FPGA resources were never exhausted, and consumed less than
25\% of the logic resources, even for complex services.

In addition to the quantitative evaluation, we note that PhD is vendor-neutral
and runtime-reconfigurable, and can be used for remote in-field debugging,
whereas standard techniques for FPGA debugging do not provide this.

\paragraph{Test setup.}
We use a host running Ubuntu server 14.04LTS, 
kernel version 4.4.0-45-generic. The host hardware is a single 3.5GHz Intel Xeon E5-2637 v4 on a SuperMicro X10-DRG-Q motherboard. An Endace 9.2SX2 DAG card (7.5ns{} time-stamping resolution) 
and a NetOptics passive-optical tap are used to intercept client-server traffic and permit independent measurement of client \& server latency. 
For the throughput tests, OSNT~\cite{antichi2014osnt} is used to control the rate
of the generated requests.

\setlength{\tabcolsep}{2pt} 
\begin{table}
\small\addtolength{\tabcolsep}{1pt}
  \begin{tabular}{lp{1cm}p{13mm}p{12mm}p{10mm}p{12mm}}
\toprule
    \bf{Artefact} & \multicolumn{2}{c}{\bf{Utilisation} (\%)}
                  & \multicolumn{3}{c}{\bf{Performance}}\\
                  & \vfill Logic & \vfill Flip-flops
                  & {\centering Duration (\#cycles)}
                  & {\centering Latency (\musec)}
                  & {\centering Queries-per-sec (KQPS)}\\
\midrule
      DNS & 100.00 & 100.00 & 57 & 1.85 & 1176 \\
      \hspace{1mm} +{\bf R} & 103.40 & 102.76 & - & 1.85 & 1176\\
      \hspace{3mm} +{\bf W} & 115.05 & 106.04 & - & 1.84 & 1176\\
      \hspace{5mm} +{\bf I} & 109.79 & 106.12 & - & 1.84 & 1176\\
\midrule
      Memcached & 100.00 & 100.00 & 64 & 2.03 & 952 \\
      \hspace{1mm} +{\bf R} & 99.17 & 100.29 & - & 2.03 & 952\\
      \hspace{3mm} +{\bf W} & 99.80 & 100.74 & - & 2.04 & 952\\
      \hspace{5mm} +{\bf I} & 100.63 & 100.69 & - & 2.03 & 952 \\
\end{tabular}

\caption{\label{tab:comparison}  Profile of utilisation and performance.
  {\bf R}ead,
  {\bf W}rite, and
  {\bf I}ncrement are instructions supported by the controller.
    Latency is indicated at the  $99^{th}$ percentile.
  The hardware generation process involves an
  optimisation step to place and route components, and on occasion this
  results in more utilisation-efficient allocations.
  The duration for extensions to DNS (clock cycles) did not change since the
  critical path of the circuit was not affected.
}
\end{table}

\begin{table}
\small\addtolength{\tabcolsep}{1pt}
  \begin{tabular}{lp{1cm}p{13mm}p{12mm}p{10mm}p{12mm}}
\toprule
    \bf{Artefact} & \multicolumn{2}{c}{\bf{Utilisation} (\%)}
                  & \multicolumn{3}{c}{\bf{Performance}}\\
                  & \vfill Logic & \vfill Flip-flops
                  & {\centering Duration (\#cycles)}
                  & {\centering Latency (\musec)}
                  & {\centering Queries-per-sec (KQPS)}\\
\midrule
      DNS+ELA & 99.74 & 100.40 & 57 & 1.83 & 1176 \\
\midrule
      DNS+2e  & 234.61 & 151.06 & 57 & 1.86 & 1176 \\
      \hspace{3mm} {\bf(Count)} & 234.46 & 151.81 & 62 & 1.94 & 1064\\
      \hspace{3mm} {\bf(Trace)} & 218.30 & 151.84 & 70 & 1.99 & 1010\\
\end{tabular}

\caption{\label{tab:more-comparison}  Utilisation and performance profile of the DNS+ELA
  against the DNS having one extension point, where
  the extension point is NOP, packet counting, or variable tracing. Latency is indicated at the  $99^{th}$ percentile.
}
\end{table}

\section{Related work}
\label{sec:relwork}
\paragraph{Portable debugging, program directing, and debugging languages.}
We were inspired by previous work on portable
debugging~\cite{Ramsey:1992:RD:143095.143112,Hood:1996:PPB:238020.238058,hanson1996machine} and program
directing~\cite{SPE:SPE4380250704}.
That work usually makes the assumption that software is compiled to run on a
general-purpose CPU however, whereas we also target reconfigurable
hardware.  The study of languages for debugging is a decades old subject~\cite{Balzer:1969:EED:1476793.1476881}
and includes sophisticated languages for high-level querying of
programs~\cite{Johnson:1977:DHL:800179.810221,hanson1993duel,winterbottom1994acid}.
Compared to that work, our work separates more starkly between the role (and
language) of the director, and that of the controller, which have separate
languages. This separation guided our modelling in~\S\ref{sec:emu:directing}.

\paragraph{Dynascope.}
Dynascope~\cite{Sosic:1992:DTP:143103.143110} provides an extremely
fine grained \emph{execution stream} of events--at the level of machine
instructions--providing a complete description of a program's runtime
behaviour.  We provide selective (and programmable) visibility by default, in
the interest of performance.  A more detailed stream could be produced if
wished.  Dynascope is generative by default (since you need to set filters to
ignore events) whereas we are generative by direction (you only get events that
you inserted code to generate).

\paragraph{High-level synthesis and runtime debugging.}
The closest work is that by \citet{McKechnie:2009:DFP:1543136.1542470} who adapt
a debugging paradigm first developed for network-on-chip devices (NoC). They provide \emph{transaction-level} granularity,
consisting of domain-specific (high-level) events--less fine grained than
Dynascope's execution stream. Compared to our
work they offer a less flexible interface, but they take a more detailed account
of different sorts of interconnects between components--formats such as
LocalLink and PLB.

Many other systems inject code to emit and observe events to aid with debugging
of complex designs, such as ``System on Chip'' designs~\cite{Lee:2015:SOF:2796316.2717310}.
SeaCucumber~\cite{Tripp2002} was the first to support source-level (HLS) debugging, both
during behavioural simulation and during hardware execution. \citet{Monson:2015:UST:2684746.2689087} take a different approach: source-code transformation
(at the HLS level) to introduce ``event observability ports'' to enable runtime
visibility of variables' values  --but note that it's not always
possible to observe an expression an interest. We took the approach of \citet{Monson:2015:UST:2684746.2689087},
relying on source transformation rather than on the HLS system.

Overhead can be reduced by being selective about what to monitor, rather than
monitoring
everything by default.  This was studied by \citet{7577371} who compared the
amount of visibility afforded by different schemes when recording events such as
reads and updates. We left the choice of what to observe to the
programmer.

At present the LegUp HLS system~\cite{6927496} appears to provide the best support for debugging.  In a contribution to that system,
\citet{Hung:2014:AFD:2597648.2566668} take a different approach to us: they use
a two-step incremental compilation, the second step of which compiles the
monitoring system by reclaiming unused FPGA resources.  This approach is less
likely to interfere with the timing behaviour of the observed circuit. Their
model is more specialised to trace-buffers, and it would be interesting to
generalise it to support our directing controller.

Testing of hardware is traditionally concentrated on the RTL description, where
tools and techniques have been developed to verify designs prior to synthesis~\cite{Foster2008,Fix2008} .
The difficulty of checking hardware entirely prior to synthesis has led to
research into the inclusion of runtime monitors in
hardware~\cite{Todman:2015:ITM:2744769.2744856}.
\citet{Curreri:2011:HSI:1972682.1972683} describe a system that translates
source-level assertions into monitors. This technique is also used to detect
hangs and possible timing overruns. We currently do not support such monitoring,
since it requires a source-level notion of time that we currently do not
provide, and ``watchdog threads'' that we currently do not include.

Finally, \citet{Camera:2005:IDE:1085130.1085145} described a hardware OS, the
Berkeley Operating system for ReProgrammable Hardware (BORPH), on which
user programs were run.
Their \emph{stitcher} extend the user program to support debugging, providing a
 rich system.
Like BORPH we require the designer to provide ``hints to the system of
what aspects of the design may need to be explored at runtime.''

\paragraph{Formalising program directors.}
Since `debugging' is such a vague term (compared to `compiling',
which has a clearer functional behaviour), its verification objective
is hard to formalise. Perhaps as a consequence of this there has been
little work on formalising and verifying debuggers, and usually
entirely theoretical~\cite{Zhu:2001:FSD:609769.609778}.
\citet{Kishon:1991:MSF:113445.113474} describe transformations over functional programs to include monitoring behaviour.
Compared to this work, our
transformations are not based on continuations (and we deliberately avoid
needing first-order functions to avoid departing too much from the conventional
hardware programming mindset).
We use an approach based on operational semantics, similar to that used by \citet{formaldebug},
but different in several ways: (i) we support a different set of
directing commands, (ii)  we don't insert commands into the bytecode
for the debugger to keep track of separations between subexpressions,
and to keep track of the path through the expression (program), (iii)
\citet{formaldebug}, introduces a language for
specifying debuggers whereas we introduce an operational language for
inspecting and changing the program's runtime state, (iv) we do not
consider the equivalence between debuggers. \citet{Sansom:1997:FBP:244795.244802} describe the profiling of functional
programs by using ``cost centres'', a paradigm to identify locations in a program at an expression-level (rather than functional-level) granularity.
Others have continued that to make it more
practical~\cite{Faddegon:2015:ADR:2737924.2737985}.
All the techniques described above rely on program transformation.
They all use functional languages as examples whereas we use a
simpler imperative language, to model the paradigm of register-level
hardware programming more directly.
Because of the weaving of debugging code into the program all these
techniques can (retrospectively) be seen as special-cases of aspect-oriented
programming, described next.

\paragraph{Aspect-oriented programming (AOP).}
AOP involves the inclusion (``weaving'') of code (called ``advice'') during compile-time or
run-time, depending on whether certain compile- or run-time conditions are
satisfied. AOP is a linchpin paradigm for tracing and monitoring
\cite{Avgustinov:2006:ATM:2172674.2172677,Hamlen:2008:AIR:1375696.1375699}
but advice can be arbitrary functions, and as a consequence they might
have undesirable effects on runtime
\cite{Avgustinov:2007:MTM:1297105.1297070}.
This is an important characteristic to our work, where we wanted to reduce
the power of added code.
\citet{DjokoDjoko:2012:APP:2108329.2108549} have categorised advice based
on the degree of influence they can have on the observable behaviour of
a program; and \citet{Dantas:2006:HA:1111037.1111071} characterise
less intrusive advice in their work on ``harmless advice''.

\section{Conclusion}
\label{sec:conc}
Having poor programming and debugging support hinders the potential of computing
architectures such as FPGAs, which are gaining importance in modern datacentres.
By using a \emph{program directing} approach we subsume several activities that
can involve the interactive runtime analyses of programs~\cite{Sosic:1992:DTP:143103.143110}.
Our language-based approach has two
extensibility benefits: (i) more efficient controllers can be implemented without
changing the source language, and (ii) third-parties could extend or customise the language of direction
commands without changing the controller that these commands compile to.

Using a program director brings security risks, since it may alter
the control-flow of a program, and this may be
exploited~\cite{Abadi:2005:CI:1102120.1102165}.  We sought to mitigate this
risk by making the presence of the debug mode very apparent, to reduce the
chances of excessive debug functionality being included in
deployment~\cite{porkexplosion}.
Dynascope was used to diagnose errors in the Dynascope compiler
itself~\cite[\S3.1]{Sosic:1992:DTP:143103.143110}, but bugs could render such a
task impossible, for reasons similar to the hereditary potential of
vulnerabilities in compilers~\cite{Thompson:1984:RTT:358198.358210}.  This
highlights the preference to have the code for the director be ``correct by
construction'', preferably automatically generated.

In work such as this we invariable come across the \emph{observer effect}~\cite{obseffect} that
monitoring code has on the monitored code. This is known to affect the visibility
of timing bugs in software~\cite{Neville-Neil:2014:OR:2661061.2661051},
but even ``invisible'' such bugs can leak important information to an adversary
\cite{Cock:2014:LME:2660267.2660294}.

The overhead due to monitoring is a very important consideration when evaluating
system measurements, and monitoring systems try to do their
utmost to reduce it~\cite{gregg2011dtrace,Anderson:2014:TTE:2592798.2592801}.
A key weakness of our prototype is that it does not use low-level (RTL-level) techniques
to reduce overhead, but we mitigate this by allowing extension points to be
specialised, e.g., these may only be a breakpoint, a watchpoint, or both, etc.
This simplifies the circuitry we get.
A more sophisticated approach would involve organising the director to
operate in a different clock domain if possible,
following~\citet{McKechnie:2009:DFP:1543136.1542470}.
Not having to sustain a large clock-distribution network leads to
power saving, since unused logic blocks don't need to be switched,
and less heat is dissipated from leakage. It also makes placement and
routing easier, which usually reduces the compilation time.

The main strength of our approach is that it provides a uniform interface for
the flexible directing of software and hardware instances of programs at
runtime. To our knowledge no other system provides this. We used an HLS system
that allows us to run the resulting code both on software and on hardware, but
the debugging methods used for either were hitherto separate.

\acks
We thank the many people who contributed to this paper. Matthew Grosvenor helped
us with evaluation ideas and reusing the QJump infrastructure. Olaf Chitil,
Paolo Costa, Klaus Gleissenthall, Tim Harris, Simon Moore, and Robert Soul\'e helped improve
the paper through their feedback.

This work has received funding from the EPSRC NaaS grant EP/K034723/1, European
Union's Horizon 2020 research and innovation programme 2014-2018 under the
SSICLOPS (grant agreement No. 644866), and the Leverhulme Trust Early Career
Fellowship ECF-2016-289.

\bibliographystyle{abbrvnat}

\appendix

\section{More program direction commands}
\label{adx:further-commands}
Program-level predicates and functions mentioned
in these definitions are given in~\S\ref{adx:metaprog}.

\subsection{Tracing}
We continue from~\S\ref{theory-rules:trace}, where the symbols we use here are
introduced (such as $X_\mathrm{i}$ and $X_\mathrm{of}$).

\begin{mathpar}
  \inferrule
    {\left(\dirTraceStartThree{X}{\isconditional}{\isresourceful}\right) \in \mathfrak{C}}
    {\directability{p}{p}{\mathfrak{C}}{\dirTraceClear{X}}
}
\end{mathpar}
\[
    \directTwo{}{}
    {\begin{array}{l}
      \fatbrack{\dirTraceClear{X}} \leadsto 0;\\
      \mathcal{D'}
    \end{array}}
\]
where
\[\begin{array}{l}
\fatbrack{\dirTraceClear{X}}
  = \\
  \qquad X_\mathrm{i}:=0; \\
  \qquad X_\mathrm{of}:=0 \\
\end{array}\]

Next we add the command to check whether the trace buffer is full.
\begin{mathpar}
  \inferrule
    {\left(\dirTraceStartThree{X}{\isconditional}{\isresourceful}\right) \in \mathfrak{C}}
    {\directability{p}{p}{\mathfrak{C}}{\dirTraceFull{X}}
}
\end{mathpar}
\[
    \directTwo{}{}
    {\begin{array}{l}
      \fatbrack{\dirTraceFull{X}} \leadsto N;\\
      \mathrm{print}(N);\\
      \mathcal{D'}
    \end{array}}
\]
where
\[\begin{array}{l}
\fatbrack{\dirTraceFull{X}}
  = X_\mathrm{of}
\end{array}\]

Finally the command to retrieve the contents of the trace buffer.
Here we first find out the current index value for the trace buffer, then
we work back and extract the buffer's contents.
\begin{mathpar}
  \inferrule
    {\left(\dirTraceStartThree{X}{\isconditional}{\isresourceful}\right) \in \mathfrak{C}}
    {\directability{p}{p}{\mathfrak{C}}{\dirTracePrint{X}}
}
\end{mathpar}
\[
    \directTwo{}{}
    {\begin{array}{l}
      X_\mathrm{i} \leadsto \mathit{cur\_idx};\\
      \text{for $i$ = $\mathit{cur\_idx}$ to $0$:}\\
      \quad X_\mathrm{a}[i] \leadsto N;\\
      \quad \mathrm{print}(N);\\
      \mathcal{D'}
    \end{array}}
\]

\subsection{Watching}
\label{theory-rules:watch}
This bears some similarity to the rule to start tracing
(in~\S\ref{theory-rules:trace}) since we inspect the variable after it's
updated, and carry out some action. When tracing, this action consists of
writing to the trace buffer. When watching, it consists of breaking.

\begin{mathpar}
\resizebox{.5 \textwidth}{!}{
  \inferrule
    {X \in \mathrm{Var}_p \\ \forall L \in X_L. \; L \not\in p \\
    \mathrm{Positions}_{p'}(X_L) = \mathrm{PostUpdate}_{p'}(X) \\ p <_{X_L} p'}
    {\directability{p}{p'}{\mathfrak{C}}
    {\dirWatch{X}{\isconditional}}
}}
\end{mathpar}
\[\directTwo{(\text{<<w>>},X,1)}{
    \begin{array}{l}
      \mathit{SP}[L \mapsto \fatbrackSP{\dirWatch{X}{\isconditional}}]\\
      \quad \text{for each $L \in X_L$}
      \end{array}}
    {\begin{array}{l}
      \mathrm{if} \; (\text{<<w>>},X,1) \in \mathcal{D'} \; \mathrm{then} \;
    \mathcal{D'}\\
      \mathrm{else} \\
      \quad \text{for each $L \in X_L$:}\\
      \qquad @L:\{\fatbrackSP{\dirWatch{X}{\isconditional}}\} \leadsto\ulcorner L\urcorner;\\
      \quad (\text{<<w>>},X,0 \mapsto 1) :\in\mathcal{D'}
    \end{array}}
\]
where
\[\begin{array}{l}
\fatbrackSP{\dirWatch{X}{\isconditional}} =
    \mathsf{conditional} \; \isconditional \; \mathtt{break}
\end{array}\]

Unwatching a variable follows a pattern we've encountered before, when stopping
tracing for instance.
\begin{mathpar}
  \inferrule
    {\left(\dirWatch{X}{\isconditional}\right) \in \mathfrak{C}
    \\ \mathrm{Positions}_{p}(X_L) = \mathrm{PostUpdate}_{p}(X)}
    {\directability{p}{p}{\mathfrak{C}}{\dirUnwatch{X}}
}
\end{mathpar}
\[
    \directTwo{}{}
    {\begin{array}{l}
      \mathrm{if} \; (\text{<<w>>},X,0) \in \mathcal{D'} \; \mathrm{then} \;
    \mathcal{D'}\\
      \mathrm{else} \\
      \quad \text{for each $L \in X_L$:}\\
      \qquad @L:\{\fatbrackSP{\dirUnwatch{X}}\} \leadsto\ulcorner L\urcorner;\\
    (\text{<<w>>},X,1 \mapsto 0):\in\mathcal{D'}
    \end{array}}
\]
where
\[\begin{array}{l}
\fatbrackSP{\dirUnwatch{X}}
  = \mathtt{continue} \\
\end{array}\]

\subsection{Profiling}
We could profile programs to count the number of writes and reads of variables,
or function calls. We show the rule for counting the number of writes;
counting variable reads and function calls are similar, differing in the positioning of
labels. That is, to count writes we position labels immediately after each
line that updates that variable--this is similar to
tracing~(\S\ref{theory-rules:trace}) and watching~(\S\ref{theory-rules:watch}).
To count variable reads we position after lines in which those variables appear.
To count function calls we position just before the first line in the function.

\newcommand{\dirCountWriteStartThree}[3]{\ensuremath{\mathsf{count} \sP \mathsf{write} \sP \mathsf{start} \sP #1 \sP #2 \sP #3}}
\newcommand{\dirCountWriteStop}[1]{\ensuremath{\mathsf{count} \sP \mathsf{write} \sP \mathsf{stop} \sP #1}}
\newcommand{\dirCountWritePrint}[1]{\ensuremath{\mathsf{count} \sP \mathsf{write} \sP \mathsf{print} \sP #1}}
\newcommand{\dirCountWriteFull}[1]{\ensuremath{\mathsf{count} \sP \mathsf{write} \sP \mathsf{full} \sP #1}}
\newcommand{\dirCountWriteClear}[1]{\ensuremath{\mathsf{count} \sP \mathsf{write} \sP \mathsf{clear} \sP #1}}

\begin{mathpar}
\resizebox{.5 \textwidth}{!}{
  \inferrule
    {X \in \mathrm{Var}_p \\ \forall L \in X_L. \; L \not\in p \\
    \mathrm{Positions}_{p'}(X_L) = \mathrm{PostUpdate}_{p'}(X) \\ p <_{X_L} p'}
    {\directability{p}{p'}{\mathfrak{C}}{\dirCountWriteStartThree{X}{\isconditional}{\isresourceful}}
}}
\end{mathpar}
\[\directTwo{(\text{<<c>>},X,1)}{
    \begin{array}{l}\mathit{C}[X_\mathrm{count}]=0, \mathit{C}[X_\mathrm{of}]=0,
      \mathit{A}[X_\mathrm{a}[\isresourceful]],\\
      \mathit{SP}[L \mapsto \fatbrackSP{\dirCountWriteStartThree{X}{\isconditional}{\isresourceful}}]\\
      \quad \text{for each $L \in X_L$}
      \end{array}}
    {\begin{array}{l}
      \mathrm{if} \; (\text{<<c>>},X,1) \in \mathcal{D'} \; \mathrm{then} \;
    \mathcal{D'}\\
      \mathrm{else} \\
      \quad \text{for each $L \in X_L$:}\\
      \qquad
      @L:\{\fatbrackSP{\dirCountWriteStartThree{X}{\isconditional}{\isresourceful}}\}\\
      \qquad\qquad\leadsto\ulcorner L\urcorner;\\
      \quad (\text{<<c>>},X,0 \mapsto 1) :\in\mathcal{D'}
    \end{array}}
\]
where
\[\begin{array}{l}
\fatbrackSP{\dirCountWriteStartThree{X}{\isconditional}{\isresourceful}} = \\
    \quad
    \mathsf{conditional} \; \isconditional \\
    \quad \left(
    \begin{array}{l}
      \mathtt{if} \; X_\mathrm{count} < \isresourceful \; \mathtt{then} \\
        \qquad \mathtt{inc} \; X_\mathrm{count}; \\
        \qquad \mathtt{continue} \\
      \mathtt{else} \\
        \qquad \mathtt{inc} \; X_\mathrm{of}; \\
        \qquad \mathtt{break}
    \end{array} \right)
\end{array}\]

Stopping the count, clearing it, printing the current count, and checking
whether the maximum value has been reached, is done in a very similar way to the
rules for tracing~(\S\ref{theory-rules:trace}).

\section{Program analyses and transformations}
\label{adx:metaprog}

\newcommand{\wedef}{\ensuremath{\stackrel{\mathit{def}}{=}}}

A position of statement $s$ in program $p$ is a vector
$\pi \in \mathbb{N}^d$ where $1 < d < \omega$.
Intuitively, the first position of the vector refers to the function in
which the statement is positioned. The second until the last positions
indicate statements (and substatements, in the case of if-then statements)
in which the statement is positioned, moving from the outermost to the
innermost containing statement. Within the innermost containing statement,
if the statement number is $i$, then the last component of the vector is
either $i + 0$ or $i + 1$, depending on whether it points to just before or
just after the statement.
\begin{defn}[Positions]
Let
\[
  p  = \overline{\vdecl} \; \overline{\fdecl} \; \return{\fname(\bar{e})}
\]
Then
\[
\Pi_p \wedef \bigcup_{0 \leq i \leq n}\left(\left\{i\right\} \times
\Pi_{\fdecl_i}\right)
\]
where $n$ is $|\overline{\fdecl}|$, the number of function declarations in $p$.

Let $\fdecl = \tau \; \fname(\bar{x})\{\overline{s}; \return{e}\}$
and $\overline{s} = s_0;\ldots;s_n$.
Then
\[
  \Pi_{\fdecl} \wedef \bigcup_{0 \leq i \leq n}\left(\left\{i\right\} \times
  \Pi_{s_i}\right)
  \cup \{n+1\}
\]
Finally,
\[
  \Pi_s \wedef \left\{
    \begin{array}{l}
      \bigcup_{0 \leq i \leq n}\left(\left\{i\right\} \times \Pi_{r_i}\right)
      \cup \{n+1\} \\ \qquad
      \text{if } s = \mathrm{if} \; e \; \mathrm{then} \; r_0;\ldots;r_n \\
      \emptyset \quad  \text{otherwise}
    \end{array}
    \right.
\]
\end{defn}

Next we define a function $\mathrm{Pos}_p$ maps a position $\pi \in \Pi_p$ in
$p$ to the corresponding statement $s$, in symbols: $\mathrm{Pos}_p(\pi) = s$.
\begin{defn}[Statement at a position]
Let
\[
  p = \overline{\vdecl} \; \overline{\fdecl} \; \return{\fname(\bar{e})}
\]
Then we define
\[
  \mathrm{Pos}_p((i, \pi)) \wedef \mathrm{FPos}_{\fdecl_i}(\pi)
\]
Let $\fdecl = \tau \; \fname(\bar{x})\{\overline{s}; \return{e}\}$
and $\overline{s} = s_0;\ldots;s_n$.
Then
\[
  \mathrm{FPos}_{\fdecl}((i, \pi)) \wedef \mathrm{SPos}_{s_i}(\pi)
\]
Finally,
\[
  \mathrm{SPos}_{s}(\pi) \wedef
    \begin{cases}
      \mathrm{SPos}_{r_i}(\pi') & \text{if} \; \pi = (i, \pi') \\
                                & \text{and} \;
        s = \mathrm{if} \; e \; \mathrm{then} \; r_0;\ldots;r_n \\
      s & \text{if} \; \pi = \emptyset
    \end{cases}
\]
\end{defn}

\begin{defn}[Program $p$ contains label $L$]
We write $L \in p$ iff $\exists \pi, S. \; \mathrm{Pos}_p(\pi) =
\mathsf{extend}(S) \; \wedge \; L \in S$.
\end{defn}

Note that a label $L$ may only be used at most once in a program.
We expect the following to hold for all programs. We do not include this as a
premise to any of the rules, to reduce clutter.
\[
  \begin{array}{l}
\forall \pi_1, \pi_2, S_1, S_2. \;
\left(\mathrm{Pos}_p(\pi_1) = \mathsf{extend}(S_1) \; \wedge \; L \in S_1\right)\\
\quad \wedge \left(\mathrm{Pos}_p(\pi_2) = \mathsf{extend}(S_2) \; \wedge \; L
    \in S_2\right) \\
\quad \longrightarrow \left(S_1 = S_2 \wedge \pi_1 = \pi_2\right)
  \end{array}
\]

We now formalise the predicate that programs $p$ and $p'$ are identical save for the
addition of $L$ in one of the extension points.
\begin{defn}[Program identity modulo label $L$]
\[
\begin{array}{l}
p <^1_L p' \wedef \forall \pi, s, S. \\
  \quad (\mathrm{Pos}_p(\pi) = s
\longrightarrow \\
\qquad (
    (s \not= \mathsf{extend}(S) \longrightarrow \mathrm{Pos}_{p'}(\pi) =
    s)
\\
\qquad \wedge \; (
 s = \mathsf{extend}(S) \longrightarrow \\
\qquad\quad \;
    \mathrm{Pos}_{p'}(\pi) = \mathsf{extend}(S')
\wedge (S' = S \vee S'\backslash S \subseteq \{L\})
)
  ))
  \\
  \quad \wedge \; (\mathrm{Pos}_{p'}(\pi) = s
\longrightarrow \\
\qquad (
    (s \not= \mathsf{extend}(S) \longrightarrow \mathrm{Pos}_{p}(\pi) =
    s)
\\
\qquad \wedge \; (
 s = \mathsf{extend}(S') \longrightarrow \\
\qquad\quad \;
    \mathrm{Pos}_{p}(\pi) = \mathsf{extend}(S)
\wedge (S' = S \vee S'\backslash S \subseteq \{L\})
)
  ))
\end{array}
\]
\end{defn}

We define a related predicate that formalises whether two programs are related
by the above predicate through an `interpolation' of labels drawn from a set.
\begin{defn}[Program identity modulo set of labels]
Let $S = \left\{L_0, \ldots, L_n\right\}$ is a set of labels.
Then,
\[
  p <_S p' \wedef
  \exists p_1,\ldots,p_{n-1}. \;
p <^1_{L_0} p_1 <^1_{L_1} p_2 <^1_{L_2} \ldots <^1_{L_n}  p'
\]
\end{defn}

\begin{defn}[Set of variables in program $p$]
Let $p = \overline{\vdecl} \; \overline{\fdecl} \; \return{\fname(\bar{e})}$
and $\overline{\vdecl} = \vdecl_0\,\ldots\,\vdecl_n$.
Then
\[
  \mathrm{Var}_p \wedef \bigcup_{0 \leq i \leq n}\{X_i\; |\; \vdecl_i = \tau_i \; X_i\}
\]
\end{defn}

\begin{defn}[Set of variables in program $p$]
\[
    \mathrm{PostUpdate}_p(X) \wedef
  \left\{ \pi + 1 \; | \; \exists E. \; \mathrm{Pos}_p(\pi) = (X := E) \right\}
\]
where if $\pi = (m_0, \ldots, m_n)$
then $\pi + 1 = (m_0, \ldots, m_n + 1)$.
\end{defn}

The positions in which a label $L$ occurs. This should be a singleton set.
\begin{defn}[Positions of label $L$]
\[\mathrm{Positions}_p(L) \wedef
  \left\{ \pi \; | \; \exists S. \;
  \mathrm{Pos}_p(\pi) = \mathsf{extend}(S) \; \wedge \;
  L \in S \right\}
\]
We lift this to work on sets of labels and overload the notation:
if $S_L$ is a set of labels then
\[
  \mathrm{Positions}_p(S_L) \wedef
\bigcup_{L \in S_L}\left(\mathrm{Positions}_p(L)\right)
\]
\end{defn}

\section{Language semantics}

\subsection{CASP machines}
\label{casp-machines:semantics}
The specification of CASP machines was given in~\S\ref{casp-machines}.

The machine's \emph{configuration} consists of the triple
$(\mathcal{S},\mathit{ia}, P)$, where $\mathit{ia} \in \{\circ, \bullet\}$
indicates whether the machine is operating in batch or interactive modes
respectively, and $\mathcal{S} = (C,A,\mathit{SP})$ indicates the machine's
memory: $C \in \left(X \rightharpoonup \mathbb{Z}\right)$ are the counters,
$A \in \left(R \rightharpoonup \mathbb{Z} \rightharpoonup \mathbb{Z}\right)$
the arrays, and
$\mathit{SP} \in \left(L \rightharpoonup P\right)$ the stored procedures.

The machine's big-step operational semantics are described using the notation
$L \vdash (\mathcal{S},\mathit{ia}, P) \Longrightarrow (\mathcal{S}',\mathit{ia}',N)$
to indicate that the machine operates in the context of a specific label $L$
to evaluate program $P$ into numeral $N$, and possibly updating the other
components of its configuration.
We use the notation $L \vdash (\mathcal{S},\mathit{ia}, P) \Downarrow N$
to abbreviate
$L \vdash (\mathcal{S},\mathit{ia}, P) \Longrightarrow (\mathcal{S},\mathit{ia},N)$
when $\mathcal{S}$ and $\mathit{ia}$ are unaffected by the evaluation.

The semantic rules are given in Figure~\ref{casp-dyn-sem}.
We use the notation \somecode{L} to denote a total injective map from labels to
numerals. Thus we can identify which label is being executed.
If in interactive mode the director sends ``break'' to the controller, the
director learns the label that broke.

Note that the ``continue'' and ``break'' commands change the state of the
$\mathit{ia}$ component of the configuration, switching it to batch (resuming
the program) and interactive respectively.

Note also that placement ($@L:\{P\}$) is only allowed during interactive mode
(i.e., $\mathit{ia} = \bullet$) since we do not want the controller to normally
run code that updates other extension points' code.

\newcommand{\caspAbbrev}[5]{\ensuremath{#1 \vdash (#2, #3, #4) \Downarrow #5}}
\newcommand{\casp}[7]{\ensuremath{#1 \vdash (#2, #3, #4)  \Longrightarrow (#5, #6, #7)}}

\newcommand{\CaspInc}[1]{\mathtt{inc}\;#1}
\newcommand{\CaspDec}[1]{\mathtt{dec}\;#1}
\newcommand{\CaspIte}[3]{\mathtt{if}\;#1\;\mathtt{then}\;#2\;\mathtt{else}\;#3}

\newcommand{\CaspSt}{\mathcal{S}}
\newcommand{\CaspIa}{\mathit{ia}}

\begin{figure*}
\begin{mathpar}
  \inferrule{}
  {\casp{L}{\CaspSt}{\CaspIa}{\CaspCont}{\CaspSt}{\circ}{\somecode{L}}}

  \inferrule{}
  {\casp{L}{\CaspSt}{\CaspIa}{\CaspBreak}{\CaspSt}{\bullet}{\somecode{L}}}

  \inferrule
    {\caspAbbrev{L}{\mathcal{S}}{\mathit{ia}}{V}{N}}
    {\caspAbbrev{L}{\mathcal{S}}{\mathit{ia}}{-V}{-N}}

  \inferrule {}{  \inferrule
    {\caspAbbrev{L}{\mathcal{S}}{\mathit{ia}}{V_1}{N_1} \\
     \caspAbbrev{L}{\mathcal{S}}{\mathit{ia}}{V_2}{N_2}
     }
    {\casp{L}{\CaspSt}{\CaspIa}{V_1 \; \mathit{op} \; V_2}{\CaspSt}{\CaspIa}{N}}
    {\quad N = \begin{cases}1 & \left(\mathit{op} = (=) \wedge N_1 = N_2\right) \vee
    \left(\mathit{op} = (<) \wedge N_1 < N_2\right) \\ -1 & \text{o/w} \end{cases}}
      }

  \inferrule {}{  \inferrule
    {\caspAbbrev{L}{\mathcal{S}}{\mathit{ia}}{U}{N}
     }
    {\casp{L}{\CaspSt}{\CaspIa}{\mathit{op} \; U}{\CaspSt[U \mapsto M]}{\CaspIa}{M}}
    {\quad M = \begin{cases}N + 1 & \mathit{op} = \mathtt{inc} \\ N - 1 & \mathit{op} = \mathtt{dec} \end{cases}}
      }

  \inferrule
    {\mathcal{S}(C, X) = N}
    {\caspAbbrev{L}{\mathcal{S}}{\mathit{ia}}{X}{N}}

  \inferrule
    {\caspAbbrev{L}{\mathcal{S}}{\mathit{ia}}{E}{N}
     }
    {\casp{L}{\CaspSt}{\CaspIa}{U := E}{\CaspSt[U \mapsto N]}{\CaspIa}{N}}

  \inferrule
    {\caspAbbrev{L}{\mathcal{S}}{\mathit{ia}}{I}{N} \\
      \mathcal{S}(R[A], N) = M}
    {\caspAbbrev{L}{\mathcal{S}}{\mathit{ia}}{A[I]}{M}}

  \inferrule {}{  \inferrule
    {\caspAbbrev{L}{\mathcal{S}}{\mathit{ia}}{E}{N} \\
     \casp{L}{\CaspSt}{\CaspIa}{P}{\CaspSt'}{\CaspIa'}{M}
     }
    {\casp{L}{\CaspSt}{\CaspIa}{\CaspIte{E}{P_1}{P_2}}{\CaspSt'}{\CaspIa'}{M}}
    {\quad P = \begin{cases}P_1 & N = 1 \\ P_2 & N = -1 \end{cases}}
      }

  \inferrule
    {\casp{L}{\CaspSt}{\CaspIa}{P_1}{\CaspSt'}{\CaspIa'}{N} \qquad \CaspIa \not= \CaspIa'}
    {\casp{L}{\CaspSt}{\CaspIa}{P_1;P_2}{\CaspSt'}{\CaspIa'}{N}}

  \inferrule
    {\casp{L}{\CaspSt}{\CaspIa}{P_1}{\CaspSt'}{\CaspIa'}{M} \qquad \CaspIa =
    \CaspIa' \qquad
     \casp{L}{\CaspSt'}{\CaspIa'}{P_2}{\CaspSt''}{\CaspIa''}{N}}
    {\casp{L}{\CaspSt}{\CaspIa}{P_1;P_2}{\CaspSt''}{\CaspIa''}{N}}

  \inferrule
    {}
    {\casp{L}{\CaspSt}{\bullet}{@L' : \{P\}}{\CaspSt[\mathit{SP}[L'] \mapsto
    P]}{\bullet}{\somecode{L'}}}

  \inferrule{}
    {\caspAbbrev{L}{\mathcal{S}}{\mathit{ia}}{N}{N}}

\end{mathpar}
  \caption{\label{casp-dyn-sem}Dynamic semantics for CASP instructions~(\S\ref{casp-machines})}
\end{figure*}

\subsection{Example language}
\label{lang:prog:semantics}
\newcommand{\konfig}[3]{\ensuremath{\langle #1, #2, #3\rangle}}
\newcommand{\step}[6]{  \ensuremath{\konfig{#1}{#2}{#3}  \longrightarrow \konfig{#4}{#5}{#6}}}
The syntax for the example language was given in~\S\ref{lang:prog}.
In this language we have a valuation function for identifiers
$R \in \left(X \rightharpoonup \mathbb{Z}\right)$, and a configuration
for the reduction semantics consists of a triple $\konfig{R}{\bar{c}}{\bar{c}}$
where $c$ is either a statement, a `return' statement, or an expression: $c ::=
s \; | \; \return{e} \; | \; e$.
$\bar{c}$ denotes a sequence of 1 or more such $c$: $c_0, \ldots, c_n$ where $n
> 0$. We define the concatenation operator $:$ such that if
$\bar{d} = d_0, \ldots, d_m$ where $m > 0$, then $\bar{c}:\bar{d} = c_0, \ldots, c_n, d_0, \ldots, d_m$.

Intuitively, $\konfig{R}{\bar{c}_1}{\bar{c}_2}$ describes a configuration where
$\bar{c}_1$ is to be evaluated ``now'', and $\bar{c}_2$ is to be executed after
$\bar{c}_1$ has been evaluated. The evaluation rules are given in Figure~\ref{proglang-dyn-sem}.

To reduce clutter in the rules, we assume two pieces implicit state, that we avoid
threading around the configurations. The first is a mapping $F$ from function names
to their bodies, and the second is the state of the CASP machine $\CaspSt$,
which includes the valuation $R$ that models the store for program variables.

\newcommand{\ifthenCommand}[2]{\ensuremath{\mathrm{if} \; #1 \; \mathrm{then} \; #2}}

We use the following evaluation context for syntactic objects in $c$
(statements, `return' statements, and expressions):
\[
\begin{array}{rcll}
  E & ::= & [] \\
    &  |  & X := E\\
    &  |  & \mathrm{if} \; E \; \mathrm{then} \; s;\overline{s} \\
    &  |  & \fname(N_0,\ldots,N_n,E,e_0\ldots,e_m)\\
    &  |  & E \; \mathit{op} \; e\\
    &  |  & N \; \mathit{op} \; E\\
    &     & \mathit{op} \in \{+, -, ==, <\} \\
\end{array}
\]

We handle `extend' by expanding it to a sequence of singleton extends, and
invoking the rule in Figure~\ref{proglang-dyn-sem-linked}.

\begin{figure*}
\begin{mathpar}
  \inferrule{}
    {\step{R}{\skipCommand}{s;\mathit{rest}}
          {R}{s}{\mathit{rest}}}

    {\step{R}{N}{s;\mathit{rest}}
          {R}{s}{\mathit{rest}}}

    {\step{R}{s;\bar{r}}{\mathit{rest}}
          {R}{s}{\bar{r}:\mathit{rest}}}

    {\step{R}{X := N}{\mathit{rest}}
          {R[X \mapsto N]}{skip}{\mathit{rest}}}

    {\step{R}{X}{\mathit{rest}}
          {R}{R(X)}{\mathit{rest}}}

  \inferrule {N \leq 0}
    {{\step{R}{\ifthenCommand{N}{\bar{s}}}{\mathit{rest}}
          {R}{skip}{\mathit{rest}}}}

  \inferrule {N > 0}
    {{\step{R}{\ifthenCommand{N}{\bar{s}}}{\mathit{rest}}
          {R}{\bar{s}}{\mathit{rest}}}}

  \inferrule
    {\step{R}{e}{\skipCommand}
          {R'}{N}{\skipCommand}}
    {\step{R}{E[e]}{\mathit{rest}}
          {R'}{E[N]}{\mathit{rest}}}

  \inferrule {}{  \inferrule
    {}
    {\step{R}{N_1 \; \mathit{op} \; N_2}{\skipCommand}
          {R}{M}{\skipCommand}}
    {\quad M = \begin{cases}
      N_1 + N_2 & \mathit{op} = +\\
      N_1 - N_2 & \mathit{op} = -\\
      1 & \mathit{op} = (==) \wedge N_1 = N_2\\
      0 & \mathit{op} = (==) \wedge N_1 \not= N_2\\
      1 & \mathit{op} = (<) \wedge N_1 < N_2\\
      0 & \mathit{op} = (<) \wedge N_1 \geq N_2\\
    \end{cases}}
      }

        \inferrule
    {(\fname, \bar{s};\return{e}) \in F \\
    \step{R}{\bar{s};\return{e}}{\skipCommand}
          {R'}{\return{N}}{\skipCommand}}
    {\step{R}{\fname(\bar{N}}{\mathit{rest}}
          {R'}{N}{\mathit{rest}}}

\end{mathpar}
  \caption{\label{proglang-dyn-sem}Dynamic semantics for the example
  language~(\S\ref{lang:prog})}
\end{figure*}

\newcommand{\roundStep}[5]{#1 \vdash (#2, #3) \curvearrowright (#4, #5)}

Rules for $\curvearrowright$ model the interactive mode between the director and
the controller. In these semantics we only see evaluation from the point of view
of the program: the program's evaluation $\longrightarrow$ yields to the
controller $\Longrightarrow$ at extension points, which either yields back to
the program (in case of continue) or else switches to interactive mode (in case
of break), leading to an interleaving between $\curvearrowright$ (obtaining a
command from the director) and $\Longrightarrow$ (executing it).
Note that the rules for $\curvearrowright$ are side-effecting, and are formalised as such
similar to streams as used by~\citet{milner1990definition}. In the rules for $\curvearrowright$,
$P$ is a command obtained from the director, and $N$ is a result sent back to
the director.

\begin{figure*}
\begin{mathpar}

  \inferrule
    {\casp{L}{\CaspSt}{\circ}{\mathit{SP}[L]}{\CaspSt'}{\circ}{N}}
    {\step{R}{\extend{L}}{s;\mathit{rest}}
          {R'}{\skipCommand}{\mathit{rest}}}

  \inferrule
    {\roundStep{L}{\CaspSt}{\bullet}{\CaspSt'}{\circ}}
    {\casp{L}{\CaspSt}{\circ}{\CaspBreak}{\CaspSt}{\circ}{\CaspCont}}

  \inferrule {}{  \inferrule
    {\casp{L}{\CaspSt}{\bullet}{P}{\CaspSt'}{\bullet}{N} \\
     \roundStep{L}{\CaspSt'}{\bullet}{\CaspSt''}{\circ}}
    {\roundStep{L}{\CaspSt}{\bullet}{\CaspSt''}{\circ}}
      }

  \inferrule {}{  \inferrule
    {\casp{L}{\CaspSt}{\bullet}{P}{\CaspSt'}{\circ}{N}}
    {\roundStep{L}{\CaspSt}{\bullet}{\CaspSt'}{\circ}}
      }

\end{mathpar}
  \caption{\label{proglang-dyn-sem-linked}Linking of the programming language
  evaluation with that of CASP controller}
\end{figure*}

\end{document}